\RequirePackage[running,mathlines]{lineno} %
\documentclass[aps,prl,nofootinbib,twocolumn,showpacs,superscriptaddress,letterpaper,amsmath,amsfonts,amssymb, preprintnumbers]{revtex4-1}  
\pdfoutput=1 

\usepackage{graphicx}
\usepackage{dcolumn}
\usepackage{bm}
\usepackage{xspace}
\usepackage{siunitx}

\usepackage{rotating}

\usepackage[colorlinks,hyperindex,breaklinks]{hyperref}
\hypersetup{linkcolor=blue,citecolor=blue,filecolor=black,urlcolor=blue} 

\usepackage{orcidlink}

\def\bulnu{\ensuremath{B \to X_u \, \ell\, \bar \nu_{\ell}}\xspace}
\def\bclnu{\ensuremath{B \to X_c \, \ell\, \bar \nu_{\ell}}\xspace}

\def\bpilnu{\ensuremath{B \to \pi \, \ell\, \bar \nu_{\ell}}\xspace}
\def\bpizlnu{\ensuremath{B^+ \to \pi^0 \, \ell^+\,\nu_{\ell}}\xspace}
\def\bpichlnu{\ensuremath{B^0 \to \pi^- \, \ell^+\,\nu_{\ell}}\xspace}
\def\brholnu{\ensuremath{B \to \rho \, \ell\, \bar\nu_{\ell}}\xspace}
\def\bomegalnu{\ensuremath{B \to \omega \, \ell \, \bar\nu_{\ell}}\xspace}
\def\betalnu{\ensuremath{B \to \eta \, \ell\, \bar \nu_{\ell}}\xspace}
\def\betaplnu{\ensuremath{B \to \eta' \, \ell\, \bar \nu_{\ell}}\xspace}
\def\bdlnu{\ensuremath{B \to D \, \ell\,\bar\nu_{\ell}}\xspace}
\def\bdslnu{\ensuremath{B \to D^* \, \ell\,\bar\nu_{\ell}}\xspace}
\def\bddslnu{\ensuremath{B \to D^{**} \, \ell\,\bar\nu_{\ell}}\xspace}

\def\Vub{\ensuremath{V_{ub}}\xspace}
\def\AbsVub{\ensuremath{|V_{ub}|}\xspace}

\newcolumntype{d}[1]{D{.}{.}{#1}}

\def\VubExclFLAG{\mbox{\ensuremath{\left|V_{ub}^{\mathrm{excl.}} \right| = (4.05 \pm 0.30 \pm 0.16 \pm 0.16)\times 10^{-3}}\xspace}}
\def\VubInclFLAG{\mbox{\ensuremath{\left|V_{ub}^{\mathrm{incl.}} \right| = (3.87 \pm 0.20 \pm 0.31 \pm 0.09)\times 10^{-3}}\xspace}}
\def\VubRatioFLAG{\mbox{\ensuremath{\left|V_{ub}^{\mathrm{excl.}} \right| / \left|V_{ub}^{\mathrm{incl.}} \right| = 1.05\pm 0.14}\xspace}}
\def\Vubcor{ \ensuremath{ \rho = 0.07}}

\def\VubExclFLAGExp{\mbox{\ensuremath{\left|V_{ub}^{\mathrm{excl.}} \right| = (3.78 \pm 0.23 \pm 0.16 \pm 0.14)\times 10^{-3}}\xspace}}
\def\VubInclFLAGExp{\mbox{\ensuremath{\left|V_{ub}^{\mathrm{incl.}} \right| = (3.88 \pm 0.20 \pm 0.31 \pm 0.09)\times 10^{-3}}\xspace}}
\def\VubRatioFLAGExp{\mbox{\ensuremath{\left|V_{ub}^{\mathrm{excl.}} \right| / \left|V_{ub}^{\mathrm{incl.}} \right| = 0.97\pm 0.12}\xspace}} 
\def\VubcorExp{\ensuremath{ \rho = 0.11}}

\def\dBF{ \ensuremath{ \Delta \mathcal{B}(B \to X_u \ell \bar \nu_\ell)& = \left( 1.39 \pm 0.14 \pm 0.22 \right) \times 10^{-3} } }

\def\BFpich{ \ensuremath{ \mathcal{B}(\overline B^0 \to \pi^+ \ell^- \bar \nu_\ell) &= \left( 1.45 \pm 0.19 \pm 0.14 \right) \times 10^{-4} } }
\def\BFcor{ \ensuremath{ \rho = 0.11}}

\def\dBFExp{ \ensuremath{ \Delta \mathcal{B}(B \to X_u \ell \bar \nu_\ell) &= \left( 1.39 \pm 0.14 \pm 0.22 \right) \times 10^{-3} } }

\def\BFpichExp{ \ensuremath{ \mathcal{B}(\overline B^0 \to \pi^+ \ell^- \bar \nu_\ell) &= \left( 1.53 \pm 0.18 \pm 0.12 \right) \times 10^{-4} } }
\def\BFcorExp{ \ensuremath{ \rho = 0.12}}

\def\BFpichHFLAV{ \ensuremath{ \mathcal{B}(\overline B^0 \to \pi^+ \ell^- \bar \nu_\ell) = \left( 1.50 \pm 0.06 \right) \times 10^{-4} } }

\begin{document}


\title{First Simultaneous Determination of Inclusive and Exclusive $\left|V_{ub}\right|$}

\noaffiliation
  \author{L.~Cao\,\orcidlink{0000-0001-8332-5668}} 
  \email{lu.cao@desy.de}
  \author{F.~Bernlochner\,\orcidlink{0000-0001-8153-2719}} 
  \email{florian.bernlochner@uni-bonn.de}
  \author{K.~Tackmann\,\orcidlink{0000-0003-3917-3761}} 
  \author{I.~Adachi\,\orcidlink{0000-0003-2287-0173}} 
  \author{H.~Aihara\,\orcidlink{0000-0002-1907-5964}} 
  \author{S.~Al~Said\,\orcidlink{0000-0002-4895-3869}} 
  \author{D.~M.~Asner\,\orcidlink{0000-0002-1586-5790}} 
  \author{H.~Atmacan\,\orcidlink{0000-0003-2435-501X}} 
  \author{T.~Aushev\,\orcidlink{0000-0002-6347-7055}} 
  \author{R.~Ayad\,\orcidlink{0000-0003-3466-9290}} 
% ...
\collaboration{The Belle Collaboration}

\begin{abstract}
The first simultaneous determination of the absolute value of the Cabibbo-Kobayashi-Maskawa matrix element \Vub using inclusive and exclusive decays is performed with the full Belle data set at the $\Upsilon(4S)$ resonance, corresponding to an integrated luminosity of 711 fb${}^{-1}$. We analyze collision events in which one $B$ meson is fully reconstructed in hadronic modes. This allows for the reconstruction of the hadronic $X_u$ system of the semileptonic $b \to u \ell \bar \nu_\ell$ decay. We separate exclusive \bpilnu decays from other inclusive \bulnu and backgrounds with a two-dimensional fit, that utilizes the number of charged pions in the $X_u$ system and the four-momentum transfer $q^2$ between the $B$ and $X_u$ system. Combining our measurement with information from lattice QCD and QCD calculations of the inclusive partial rate as well as external experimental information on the shape of the \bpilnu form factor, we determine \VubExclFLAGExp\ and \VubInclFLAGExp, respectively, with the uncertainties being the statistical error, systematic errors, and theory errors. The ratio of $\left|V_{ub}^{\mathrm{excl.}} \right| / \left|V_{ub}^{\mathrm{incl.}} \right| = 0.97 \pm 0.12$ is compatible with unity. 

\end{abstract}


\preprint{Belle Preprint 2023-04, KEK Preprint 2022-53}

\maketitle

In this letter we report the first simultaneous determination of the absolute value of the Cabibbo-Kobayashi-Maskawa (CKM) matrix element \Vub\ using inclusive and exclusive decays. Exclusive determinations of \AbsVub\ focus on reconstructing explicit final states such as $B \to \pi \ell \bar \nu_\ell$~\cite{HFLAV:2022pwe}, $\Lambda_b \to p \mu \bar \nu_\mu$~\cite{Aaij:2015bfa}, or $B_s \to K \, \mu \bar \nu_\mu$~\cite{LHCb:2020ist}, whereas inclusive determinations study $B$ meson decays undergoing $b \to u \ell \bar \nu_\ell$ transitions and are indiscriminate of the $u \to X_u$ hadronization process. 
 The world averages of either method are only marginally compatible~\cite{HFLAV:2022pwe},
\begin{linenomath*}
\begin{align}
 |V_{ub}^{\mathrm{excl.}}| & =  \left(3.51 \pm 0.12 \right) \times 10^{-3} \, , \\
 |V_{ub}^{\mathrm{incl.}}| & = \left( 4.19 \pm 0.16 \right) \times 10^{-3} \, ,
\end{align}
\end{linenomath*}
with a ratio of $ |V_{ub}^{\mathrm{excl.}}| /  |V_{ub}^{\mathrm{incl.}}|  = 0.84 \pm 0.04$, which deviates 3.7 standard deviations from unity. The underlying reason for this tension is unknown. New physics explanations are challenging (see e.g. Refs.~\cite{Crivellin:2009sd,Enomoto:2014cta,Bernlochner:2014ova,Umeeda:2022dmz}), leading to some to speculate the existence of until now unaccounted systematic effects~\cite{pdg_Vxb:2020}. This motivates the simultaneous determination in a single analysis, in which $B \to \pi \ell \bar \nu_\ell$ and the \bulnu\ rates can be simultaneously extracted and systematic effects can be correlated.

The presented measurement of inclusive and exclusive $b \to u \ell \bar \nu_\ell$ decays uses the same collision events and a similar analysis strategy as Refs.~\cite{Cao:2021xqf, Belle:2021ymg}. Charmless semileptonic decays are reconstructed by relying on the complete reconstruction of the second $B$ meson in the $e^+ e^- \to \Upsilon(4S) \to B \bar B$ process. This approach allows for the direct reconstruction of the $X_u$ system of the \bulnu process. Specifically, the four-momentum transfer squared, $q^2 = \left( p_B - p_{X_u} \right)^2$, and the number of charged pion candidates of the $X_u$ system, $N_{\pi^\pm}$, can be reconstructed. This allows for the separation of \bpizlnu and \bpichlnu from other \bulnu decays. The main background in the measurement stems from the much more abundant \bclnu decays and a multivariate suppression strategy is used to reduce this and other background processes. Charge conjugation is implied throughout. The inclusive \bulnu branching fraction is defined as the average branching fraction of $B^+$ and $B^0$ meson decays. Furthermore, we denote $\ell = e, \mu$, and use natural units: $\hbar = c = 1$.

We analyze \mbox{$(772 \pm 10) \times 10^6$} $B$ meson pairs recorded at the $\Upsilon(4S)$ resonance energy and $\SI{79}{fb^{-1}}$ of collision events recorded $\SI{60}{MeV}$ below the $\Upsilon(4S)$ peak. Both data sets were recorded at the KEKB $e^+ e^-$ collider~\cite{KEKB} by the Belle detector. Belle is a large-solid-angle magnetic spectrometer. A detailed description of its performance and subdetectors can be found in Ref.~\citep{Abashian:2000cg}. The particle identification and selection criteria are the same as in Ref.~\cite{Cao:2021xqf}. 

Monte Carlo (MC) samples of $B$ meson decays and continuum processes ($e^+ e^- \to q \bar q$ with $q = u,d,s,c$) are simulated using the \texttt{EvtGen} generator~\citep{EvtGen}. The normalization of continuum events is calibrated with the measured off-resonance data. A detailed description of all samples and decay models is given in Ref.~\cite{Cao:2021xqf}. The simulated samples are used for background subtraction and to correct for detector resolution, selection, and acceptance effects. The used sample sizes correspond to approximately ten and five times, respectively, the Belle collision data for the $B$ meson production and continuum processes. 

Semileptonic \bulnu decays are simulated as a mixture of specific exclusive modes and nonresonant contributions using a ``hybrid'' approach~\cite{hybrid,Prim:2019gtj,markus_prim_2020_3965699}: the triple differential rate of inclusive and exclusive predictions are combined such that the partial rates of the inclusive prediction are recovered. This is achieved by assigning weights to the inclusive contribution as a function of the generator-level $q^2$, $E_\ell^B$, and $M_X$. Here $E_\ell^B$ and $M_X$ denote the energy of the lepton in the signal $B$ rest frame and the invariant mass of the $X_u$ system produced in the \bulnu decay. For the inclusive contribution, we use two different calculations: the De Fazio and Neubert (DFN) model~\cite{DeFazio:1999ptt} (with $m_{b}^{\text{KN}} = (4.66 \pm 0.04)\,\mathrm{GeV}$, $a^{\text{KN}} = 1.3 \pm 0.5$) and the Bosch-Lange-Neubert-Paz (BLNP) model~\cite{Lange:2005yw} (with $m_{b}^{\mathrm{SF}} = 4.61\,\mathrm{GeV}$, $\mu_{\pi}^{2\, \text{SF}} = 0.20 \,\mathrm{GeV}^2$). The difference between the two models is treated as a systematic uncertainty. The simulated inclusive \bulnu events are hadronized with the JETSET algorithm~\cite{SJOSTRAND199474} into final states with two or more mesons. We study two different tunes of the fragmentation parameters and assign their difference as a systematic uncertainty. The exclusive contributions are modeled as follows: \bpilnu decays are modeled using the Bourrely-Caprini-Lellouch (BCL) form factor parameterization~\citep{Bourrely:2008za}; \brholnu and \bomegalnu decays are modeled using the Bharucha-Straub-Zwicky (BSZ) form factors~\citep{Bharucha:2015bzk} from the fit of Ref.~\citep{Bernlochner:2021rel} to light-cone sum rule (LCSR) predictions~\citep{Bharucha:2015bzk} and the measurements of Refs.~\cite{Sibidanov:2013rkk,Lees:2012mq,delAmoSanchez:2010af}; \betalnu and \betaplnu are modeled using pole form factors obtained from fits to LCSR~\citep{Duplancic:2015zna}. For the  branching fractions the world averages from Ref.~\citep{pdg:2020} are used. 

Semileptonic \bclnu\ decays are dominated by \bdlnu\ and \bdslnu\ decays. We simulate them with the form factors of Refs.~\cite{Boyd:1994tt,Grinstein:2017nlq,Bigi:2017njr} and values determined by the measurements of Refs.~\cite{Glattauer:2015teq,Waheed:2018djm}. Other \bclnu\ decays are simulated as a mixture of resonant and nonresonant modes, using the parameterization of Ref.~\cite{Bernlochner:2016bci} for the modeling of \bddslnu form factors. The known difference between inclusive and the sum of measured exclusive \bclnu decays is simulated with $B \to D^{(*)} \, \eta  \, \ell^+  \nu_\ell$ decays.

We reconstruct $e^+ e^-$ collision events with the multivariate tagging algorithm of Ref.~\cite{Feindt:2011mr}. The algorithm uses a hierarchical approach utilizing neural networks to fully reconstruct one of the two $B$ mesons in hadronic final states (labeled as $B_{\mathrm{tag}}$). The $B_{\mathrm{tag}}$ reconstruction efficiency is calibrated using \bclnu decays following the prescription outlined in~\cite{Cao:2021xqf}. The identified final state particles forming the $B_{\mathrm{tag}}$ are masked and $b \to u \ell \bar \nu_\ell$ signal candidates are reconstructed by identifying an electron or muon candidate in the events, requiring \mbox{$E_\ell^B = |\bold{p}_{\ell}^B| > 1 \, \mathrm{GeV}$} as measured in the signal $B$ rest frame. To reject background from the much more abundant \bclnu decays, eleven distinguishing features are combined into a single discriminant using boosted decision trees (BDTs) and utilizing the implementation of Ref.~\cite{Chen:2016:XST:2939672.2939785}. The most discriminating training features are the reconstructed neutrino mass, $M_{\mathrm{miss}}^2$, the vertex fit probability of the decay vertex between the hadronic system $X $ and the signal lepton $\ell$, and the number of identified $K^\pm$ and $K_{S}^{0}$ in the $X$ system. Same as in~\cite{Cao:2021xqf}, we select a working point that corresponds to a signal efficiency of 18.5\%, which rejects 98.7\% of \bclnu\ decays, defined with respect to all events after the $B_{\mathrm{tag}}$ selection. To test the modeling of \bclnu and other backgrounds in the extraction variables, $q^2$ and $N_{\pi^\pm}$, we also utilize the events failing the BDT selection and find good agreement~\cite{supplemental}. We further separate events by the reconstructed $M_X$, categorizing $M_X < 1.7 \, \mathrm{GeV}$ into five $q^2$ bins ranging in $[0, 26.4]\, \mathrm{GeV}^{2}$ as a function of the $N_{\pi^\pm}$ multiplicity for the interval of $[0,1,2,\geq 3]$. Events with $M_X \geq 1.7 \, \mathrm{GeV}$ are analyzed only in bins of $N_{\pi^\pm}$ as they are dominated by background. To enhance the \bpilnu\ purity in the low-$M_{X}$ $N_{\pi^\pm} = 0$ and $N_{\pi^\pm} = 1$ events, we apply a selection on the thrust of 0.92 and 0.85, respectively. It is defined by $\max_{|\bold{n}| = 1} \left( \sum_i | \bold{p_i} \cdot \bold{n}| / \sum_i  | \bold{p_i} |  \right)$, when summing over the neutral and charged constituents of the reconstructed $X$ system in the center of mass frame. For \bpilnu\ events, we expect a more collimated $X_u$ system than for \bclnu and other \bulnu processes, resulting in a higher thrust value.

The $q^2:N_{\pi^\pm}$ bins and the $M_X \geq 1.7 \, \mathrm{GeV}$ $N_{\pi^\pm}$ distribution are analyzed using a simultaneous likelihood fit, which incorporates floating parameters for the modeling of the \bpilnu form factor, the binned templates, and systematic uncertainties as nuisance parameters. Specifically, the shape of \bpilnu template is linked to the form factors by correcting the efficiency and acceptance effects. The fit components we probe are the normalizations of \bpilnu decays, other \bulnu signal decays, and of background events dominated by \bclnu decays. The $f_+$ and $f_0$ form factors describing the \bpilnu decay dynamics are parameterized with expansion coefficients $a_n^+ $ and $a_n^0$ using  the BCL expansion,
\begin{linenomath*}
\begin{align} \label{eq:bcl}
 f_+(q^2) & =  \frac{1}{ 1 - q^2 / m_{B^*}^2} \sum_{n = 0}^{N^+ - 1}\, a_n^+ \left[ z^n - (-1)^{n-N^+} \frac{n}{N^+} \, z^{N^+} \right] \, , \nonumber \\
 f_0(q^2) & =  \sum_{n = 0}^{N^0 - 1}\, a_n^0 \, z^n \, ,
\end{align}
\end{linenomath*}
at expansion order $N^+= N^0 =3$ in the conformal variable $z = z(q^2)$~\citep{Bourrely:2008za,FLAG:2021npn}, and $a^0_2$ is expressed by the remaining coefficients to keep the kinematical constraint $f_{+}(0) = f_{0}(0)$. We constrain the expansion coefficients to the lattice QCD (LQCD) values of Ref.~\cite{FLAG:2021npn}, combining LQCD calculations from several groups~\cite{FermilabLattice:2015mwy,Flynn:2015mha}. Note that the measured distributions have no sensitivity for $f_0$ and we thus neglect its effects in the decay rate. The inclusion of the $f_0$ expansion coefficients, however, reduces uncertainties on the \bpilnu rate through the correlation to the $f_+$ shape. In order to utilize the full experimental knowledge of the $B \to \pi$ form factors to date, we constrain its shape to the combined lattice QCD and experimental information of Refs.~\cite{BaBar:2010efp,BaBar:2012thb,Belle:2010hep,Belle:2013hlo}. The fit scenario with only lattice QCD constraints is studied for a standalone comparison with other experimental results.

We consider additive and multiplicative systematic uncertainties in the likelihood fit by adding bin-wise nuisance parameters for each template. The parameters are constrained to a multinormal Gaussian distribution with a covariance reflecting the sum of all considered systematic effects, and the correlation structure between templates from common sources is taken into account. This includes detector and reconstruction related uncertainties, such as the tracking efficiency for low and high momentum tracks, particle identification efficiency uncertainties, and the calibration of the $B_{\mathrm{tag}}$ reconstruction efficiency. We further consider uncertainties on the \bulnu and \bclnu shapes from the form factors, non-perturbative parameters, and their compositions. 
The $u \to X_u$ fragmentation uncertainties are evaluated by changing the default Belle tune of fragmentation parameters to the values used in Ref.~\cite{LHCb:2014wmv}. We further vary the $s \bar s$-production rate $\gamma_s = 0.30 \pm 0.09$, spanning the range of Refs.~\cite{Althoff:1984iz,Bartel:1983qp}. The largest uncertainties on the exclusive branching fraction measurements are from the calibration of the tagging efficiency ($\pm 4.1\%$) and the \bulnu modeling ($\pm 3.5\%$). The largest uncertainties on the inclusive branching fraction measurement are from the \bulnu  ($\pm 10.9\%$) modeling and the $u \to X_u$ fragmentation ($\pm 5.3\%$). The uncertainties of the modeling of the \bclnu background are $\pm 1.2\%$ and $\pm 2.8\%$ for the \bpilnu and \bulnu branching fractions, respectively.

\begin{figure}
	\centering
		 \includegraphics[width=1.0\linewidth]{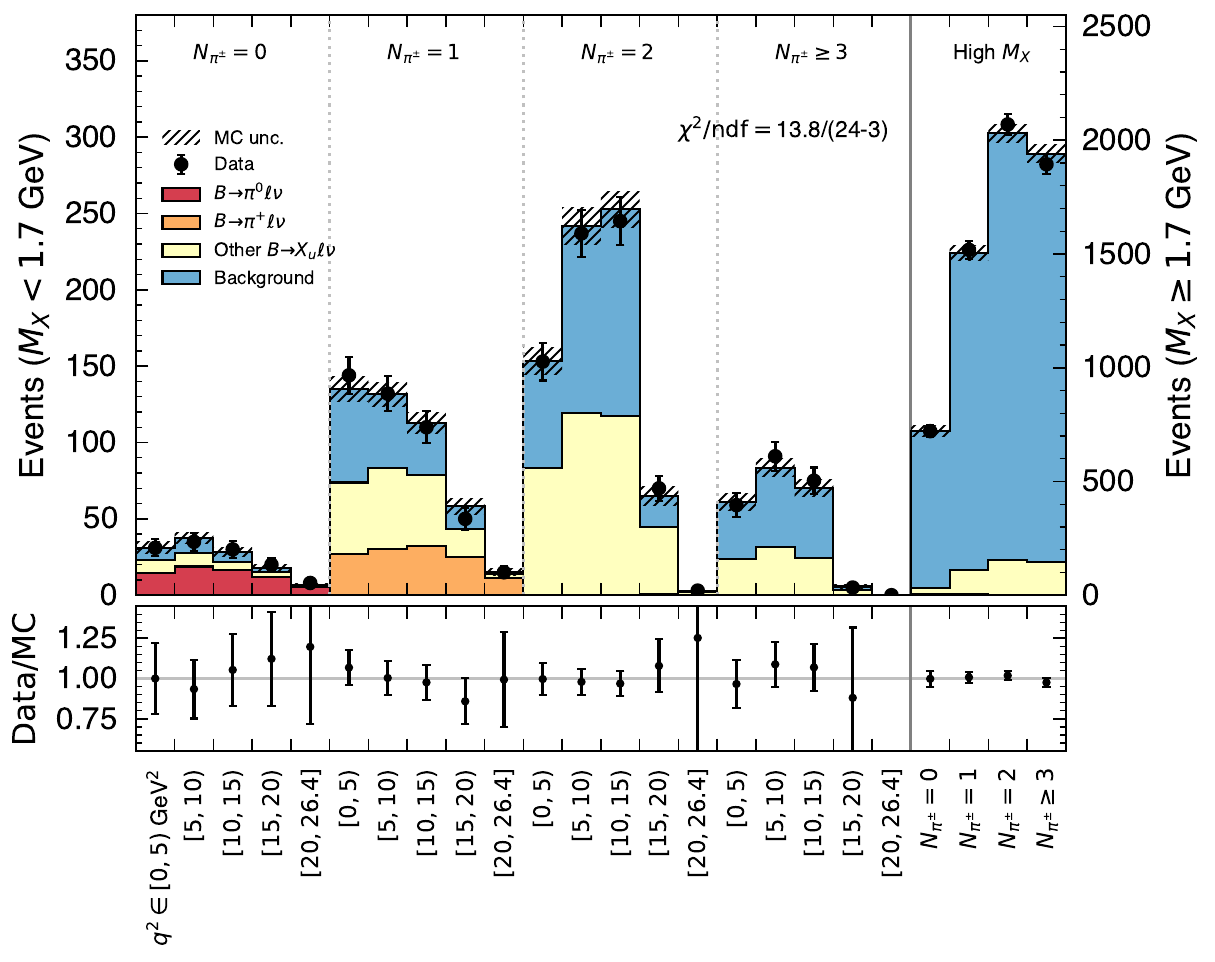} 
	\caption{\label{fig:sig} 
	The $q^2:N_{\pi^\pm}$ spectrum after the 2D fit is shown for the scenario that only uses LQCD information. The uncertainties incorporate all postfit uncertainties discussed in the text. 
	} 
\end{figure}

\begin{figure}[ht!]
	\centering
		 \includegraphics[width=0.95\linewidth]{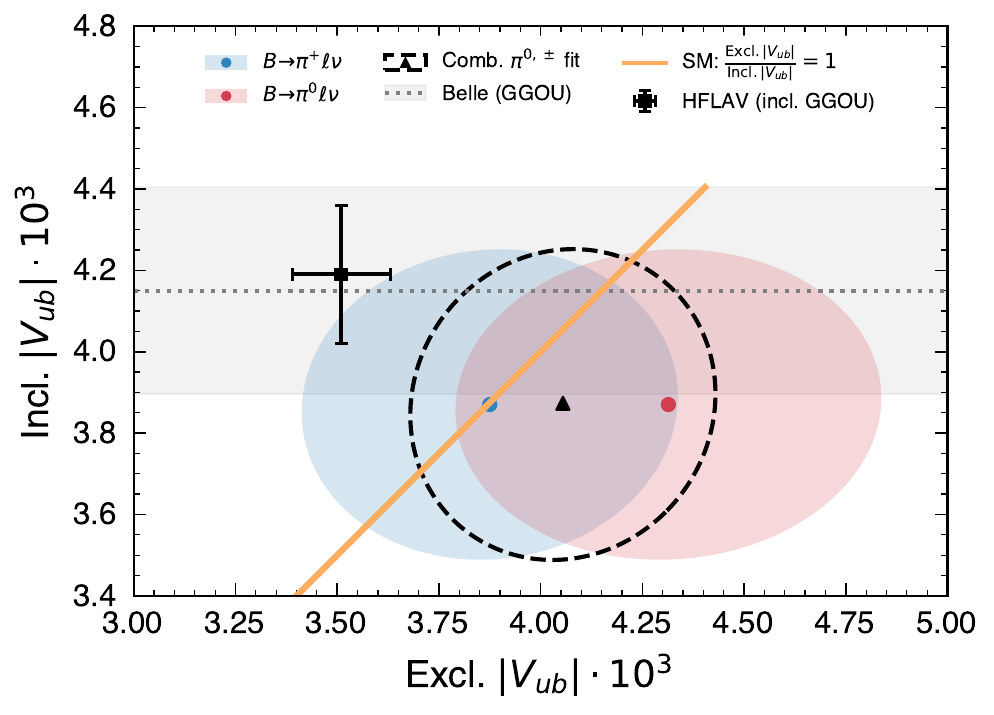}   \\
		 \includegraphics[width=0.95\linewidth]{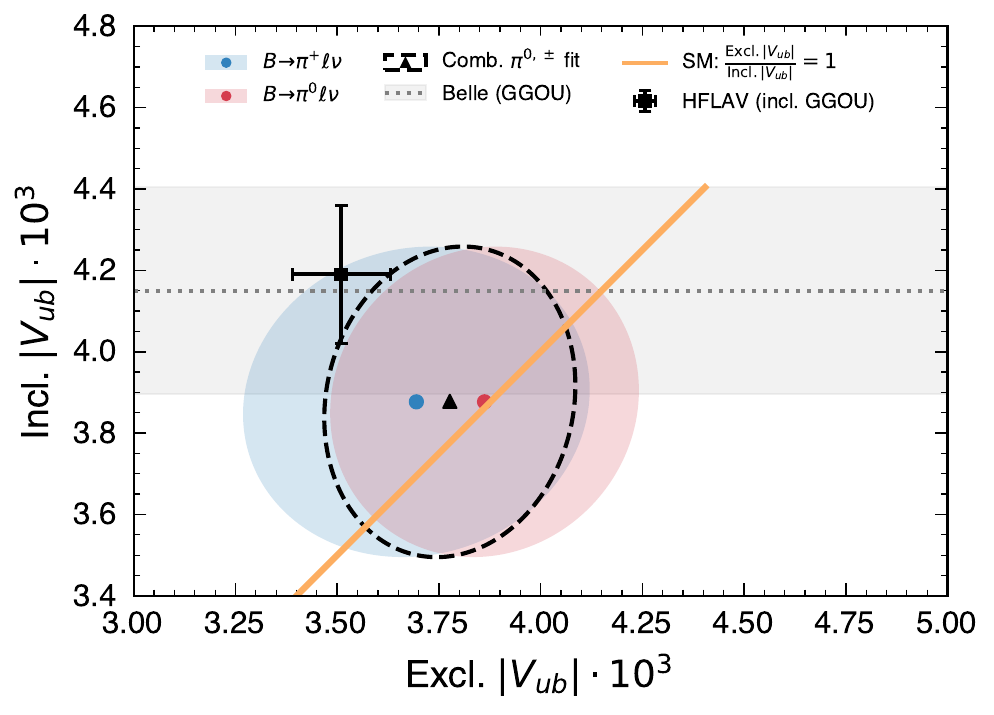} 		 
	\caption{ 
	The \AbsVub\ values obtained with the fits using (top) LQCD or (bottom) LQCD and experimental constraints for the $\overline B^0 \to \pi^+ \ell^- \bar \nu_\ell$ form factor are shown. The inclusive \AbsVub value is based on the decay rate from the GGOU calculation. The values obtained from the previous Belle measurement ~\cite{Cao:2021xqf} (grey band) and the world averages from Ref.~\cite{HFLAV:2022pwe} (black marker) are also shown. The shown ellipses correspond to 39.3\% confidence levels ($\Delta \chi^2 = 1$).
	}
\label{fig:npi-vub-FLAG} 	
\end{figure}

Figure~\ref{fig:sig} shows the $q^2:N_{\pi^\pm}$ distribution of the signal region after the fit and with only using LQCD information: \bpizlnu and \bpichlnu events are aggregated in the $N_{\pi^+} = 0$ and $N_{\pi^+} = 1$ bins, respectively, whereas contributions from other \bulnu\ processes are in all multiplicity bins. The high $M_X$ bins constrain the \bclnu and other background contributions. We use the isospin relation and $B^0/B^+$ lifetime ratio to link the yields of \bpizlnu and \bpichlnu. The fit has a $\chi^2$ of $13.8$ with $21$ degrees of freedom, corresponding to a p-value of $88\%$. The measured \bpizlnu and \bpichlnu yields are corrected for efficiency effects to determine the corresponding branching fractions $\mathcal{B}$. The measured inclusive yield is calculated from the sum of \bpizlnu, \bpichlnu, and other \bulnu events and unfolded to correspond to a partial branching fraction $\Delta\mathcal{B}$ with $E_\ell^B > 1.0 \, \mathrm{GeV}$, also correcting for the effect of final state radiation photons. We find
\begin{linenomath*}
\begin{align}
  \BFpich \, , \\   
    \dBF \, ,
\end{align}
\end{linenomath*}
with the errors denoting statistical and systematic uncertainties and we used the isospin relation between $B^- \to \pi^0 \ell^- \bar \nu_\ell$ and $\overline B^0 \to \pi^+ \ell^- \bar \nu_\ell$ to link both branching fractions. The recovered branching fraction for \mbox{$\overline B^0 \to \pi^+ \ell^- \bar \nu_\ell$} is compatible with the world average of \mbox{\BFpichHFLAV}~\cite{HFLAV:2022pwe}. The correlation between the exclusive and inclusive branching fractions is \BFcor. Using calculations for the inclusive partial rate and the fitted form factor parameters, we can determine values for \AbsVub. As our baseline we use the GGOU~\cite{GGOU} calculation for the inclusive partial rate with $E_\ell^B > 1.0 \, \mathrm{GeV}$ (\mbox{$\Delta\Gamma /  |V_{ub}|^2  = 58.5 \pm 2.7$ ps$^{-1}$}), but other calculations result in similar values for inclusive \AbsVub. We find
\begin{linenomath*}
\begin{align}
\VubExclFLAG \, , \\
\VubInclFLAG \, , 
\end{align}
\end{linenomath*}
for exclusive and inclusive \AbsVub with the uncertainties denoting the statistical error, systematic error, and error from theory (either from LQCD or the inclusive calculation). The correlation between the exclusive and inclusive \AbsVub is \Vubcor. The determined value for inclusive \AbsVub is compatible with the determination of Ref.~\cite{Cao:2021xqf}. For the ratio of inclusive and exclusive \Vub values, we find
\begin{linenomath*}
\begin{align}
\VubRatioFLAG \, ,
\end{align} 
\end{linenomath*}
which is compatible with the SM expectation of unity. The value is higher and compatible with the current world average of $ |V_{ub}^{\mathrm{excl.}}| /  |V_{ub}^{\mathrm{incl.}}|  = 0.84 \pm 0.04$~\cite{HFLAV:2022pwe} within 1.5 standard deviations. Fig.~\ref{fig:npi-vub-FLAG} (top) compares the measured individual values with the SM expectation and the current world average. We also test what happens if we relax the isospin relation between $B^- \to \pi^0 \ell^- \bar \nu_\ell$ (red ellipse) and $\overline B^0 \to \pi^+ \ell^- \bar \nu_\ell$ (blue) branching fractions and find compatible results for exclusive and inclusive \AbsVub, as well as for the exclusive \AbsVub values.

\begin{figure}
	\centering
		 \includegraphics[width=1\linewidth]{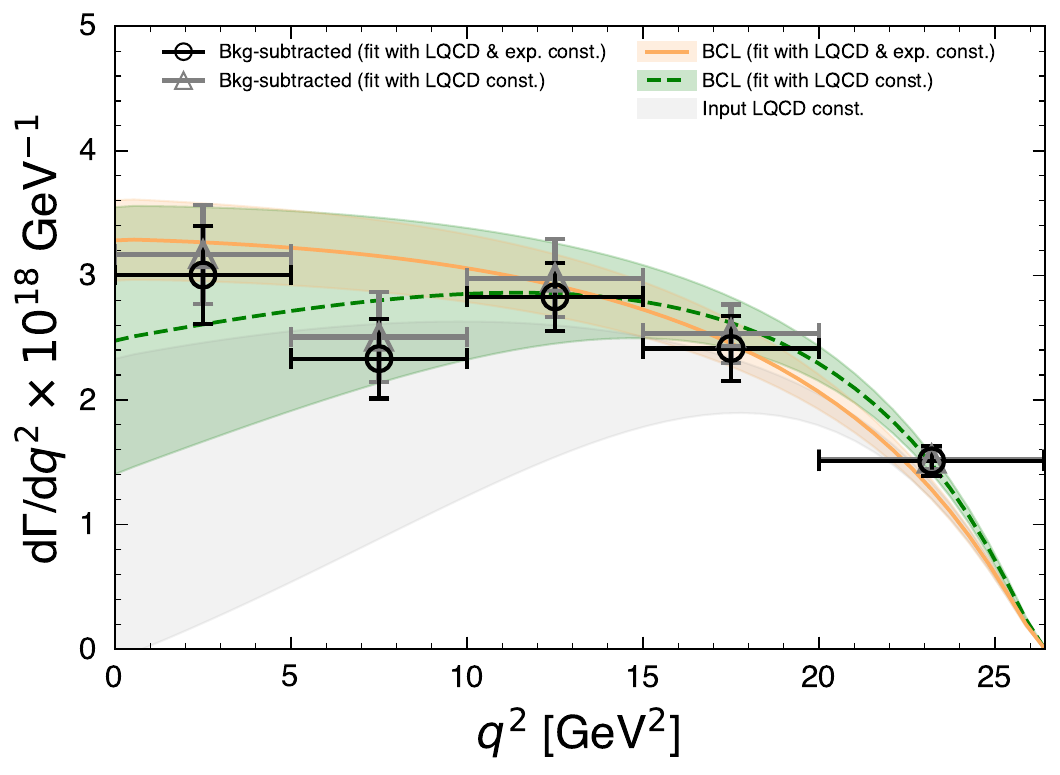}  
	\caption{ 
	The $q^{2}$ spectra of $\overline B^0 \to \pi^+ \ell^- \bar \nu_\ell$ obtained from the fit of the combined LQCD and experimental information (orange, solid) and from the fit to LQCD only (green, dashed) are shown. The data points are the background subtracted post-fit distributions, corrected for resolution and efficiency effects and averaged over both isospin modes. In addition, the LQCD pre-fit prediction of~\cite{FLAG:2021npn} for the $\overline B^0 \to \pi^+ \ell^- \bar \nu_\ell$ form factor is shown (grey). 
	}
\label{fig:BCL-FLAG} 	
\end{figure}

In the nominal result, we utilize the full theoretical and experimental knowledge of the \bpilnu form factor, combining shape information from the measured $q^2$ spectrum with LQCD predictions, as provided by Ref.~\cite{FLAG:2021npn}. The determined (partial) branching fractions in this scenario are
\begin{linenomath*}
\begin{align}
  \BFpichExp \, , \\
    \dBFExp \, ,
\end{align}
\end{linenomath*}
with a correlation of \BFcorExp\ between inclusive and exclusive branching fractions and assuming isospin relation. This fit leads to a more precise value of \AbsVub\ from \bpilnu and we find with the same inclusive calculation
\begin{linenomath*}
\begin{align}
   \VubExclFLAGExp \, , \\ 
   \VubInclFLAGExp \, ,  
\end{align}
\end{linenomath*}
with a correlation \VubcorExp\ and a ratio of 
\begin{align}
\VubRatioFLAGExp \, ,
\end{align}
compatible with the world average within 1.2 standard deviations. Fig.~\ref{fig:npi-vub-FLAG} (bottom) compares the obtained values and we also find good agreement between the isospin conjugate exclusive values of \AbsVub. Figure~\ref{fig:BCL-FLAG} compares the fitted $q^2$ spectra of the differential rate of $\overline B^0 \to \pi^+ \ell^- \bar \nu_\ell$ for both fit scenarios as well as for the LQCD input~\cite{FLAG:2021npn}. The inclusion of the full experimental and theoretical knowledge leads to a higher rate at low $q^2$.

In summary, we presented the first simultaneous determination of inclusive and exclusive \AbsVub\ within a single analysis. In the ratio of both \AbsVub\ values many systematic uncertainties such as the tagging calibration or the lepton identification uncertainties cancel and one can directly test the SM expectation of unity. We recover ratios that are compatible with this expectation, but 1.5 standard deviations higher than the ratio of the current world averages of inclusive and exclusive \AbsVub. This tension is reduced to 1.2 standard deviations when including the constraint based on the full theoretical and experimental knowledge of the \bpilnu form factor shape. We average our inclusive and exclusive values from both approaches using LQCD or LQCD and additional experimental information and find,
\begin{linenomath*}
\begin{align}
 |V_{ub}| & =  (3.96 \pm 0.27)\times 10^{-3} \, , \quad (\mathrm{LQCD}) \\
 |V_{ub}| & =  (3.84 \pm 0.26)\times 10^{-3}\, ,   \quad  (\mathrm{LQCD+exp.})
\end{align}
\end{linenomath*}
respectively. These values can be compared with the expectation from CKM unitarity of Ref.~\cite{CKMfitter2021} of \mbox{$|V_{ub}^{\mathrm{CKM}}| = (3.64 \pm 0.07)\times 10^{-3}$} and are compatible within 1.2 and 0.8 standard deviations, respectively. The applied approach of simultaneously fitting $q^2$ and the number of charged pions in the $X_u$ system will benefit from the large anticipated data set of Belle~II. Additional fit scenarios and inclusive \AbsVub values from other theory calculations of the partial rate are provided in the supplemental material~\cite{supplemental}.

This work, based on data collected using the Belle detector, which was
operated until June 2010, was supported by 
the Ministry of Education, Culture, Sports, Science, and
Technology (MEXT) of Japan, the Japan Society for the 
Promotion of Science (JSPS), and the Tau-Lepton Physics 
Research Center of Nagoya University; 
the German Research Foundation (DFG) Emmy-Noether Grant No. BE 6075/1-1; 
the Helmholtz W2/W3-116 grant;
the Australian Research Council including grants
DP180102629, 
DP170102389, 
DP170102204, 
DE220100462, 
DP150103061, 
FT130100303; 
Austrian Federal Ministry of Education, Science and Research (FWF) and
FWF Austrian Science Fund No.~P~31361-N36;
the National Natural Science Foundation of China under Contracts
No.~11675166,  
No.~11705209;  
No.~11975076;  
No.~12135005;  
No.~12175041;  
No.~12161141008; 
Key Research Program of Frontier Sciences, Chinese Academy of Sciences (CAS), Grant No.~QYZDJ-SSW-SLH011; 
Project ZR2022JQ02 supported by Shandong Provincial Natural Science Foundation;
the Ministry of Education, Youth and Sports of the Czech
Republic under Contract No.~LTT17020;
the Czech Science Foundation Grant No. 22-18469S;
Horizon 2020 ERC Advanced Grant No.~884719 and ERC Starting Grant No.~947006 ``InterLeptons'' (European Union);
the Carl Zeiss Foundation, the Deutsche Forschungsgemeinschaft, the
Excellence Cluster Universe, and the VolkswagenStiftung;
the Department of Atomic Energy (Project Identification No. RTI 4002) and the Department of Science and Technology of India; 
the Istituto Nazionale di Fisica Nucleare of Italy; 
National Research Foundation (NRF) of Korea Grant
Nos.~2016R1\-D1A1B\-02012900, 2018R1\-A2B\-3003643,
2018R1\-A6A1A\-06024970, RS\-2022\-00197659,
2019R1\-I1A3A\-01058933, 2021R1\-A6A1A\-03043957,
2021R1\-F1A\-1060423, 2021R1\-F1A\-1064008, 2022R1\-A2C\-1003993;
Radiation Science Research Institute, Foreign Large-size Research Facility Application Supporting project, the Global Science Experimental Data Hub Center of the Korea Institute of Science and Technology Information and KREONET/GLORIAD;
the Polish Ministry of Science and Higher Education and 
the National Science Center;
the Ministry of Science and Higher Education of the Russian Federation, Agreement 14.W03.31.0026, 
and the HSE University Basic Research Program, Moscow; 
University of Tabuk research grants
S-1440-0321, S-0256-1438, and S-0280-1439 (Saudi Arabia);
the Slovenian Research Agency Grant Nos. J1-9124 and P1-0135;
Ikerbasque, Basque Foundation for Science, Spain;
the Swiss National Science Foundation; 
the Ministry of Education and the Ministry of Science and Technology of Taiwan;
and the United States Department of Energy and the National Science Foundation.
These acknowledgements are not to be interpreted as an endorsement of any
statement made by any of our institutes, funding agencies, governments, or
their representatives.
We thank the KEKB group for the excellent operation of the
accelerator; the KEK cryogenics group for the efficient
operation of the solenoid; and the KEK computer group and the Pacific Northwest National
Laboratory (PNNL) Environmental Molecular Sciences Laboratory (EMSL)
computing group for strong computing support; and the National
Institute of Informatics, and Science Information NETwork 6 (SINET6) for
valuable network support.

We are indebted to Alexander Ermakov for his pioneering work on the subject. We thank Frank Tackmann, Zoltan Ligeti, and Dean Robinson for discussions about the content of this manuscript. 

\bibliographystyle{apsrev4-1}
\bibliography{InclExclVub}

\clearpage
\newpage
\onecolumngrid
\section*{Supplemental Material} 
\setcounter{page}{1}

\subsection{Determinations with alternative inclusive calculations for the partial rate}

\begin{figure}[ht!]
	\centering
		 \includegraphics[width=0.49\linewidth]{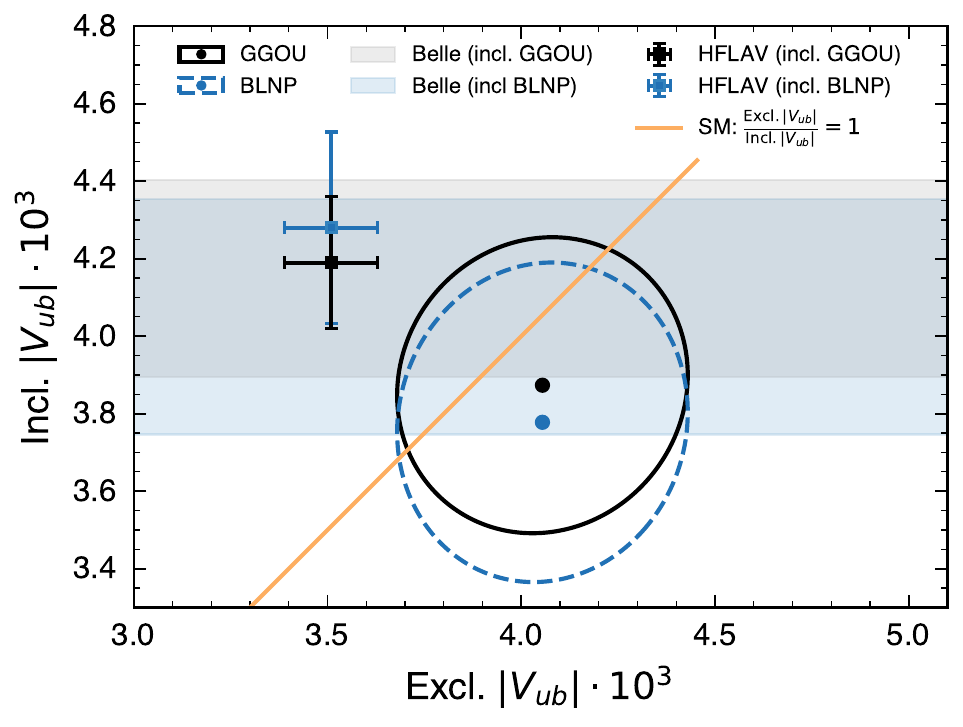} 
		  \includegraphics[width=0.49\linewidth]{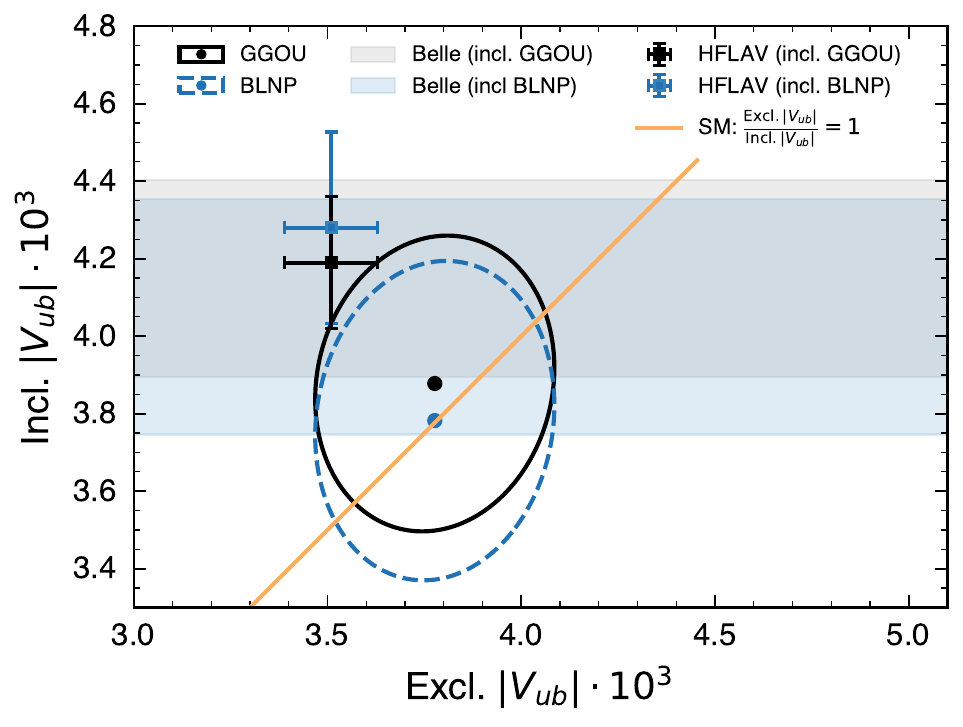}  \\
            \includegraphics[width=0.49\linewidth]{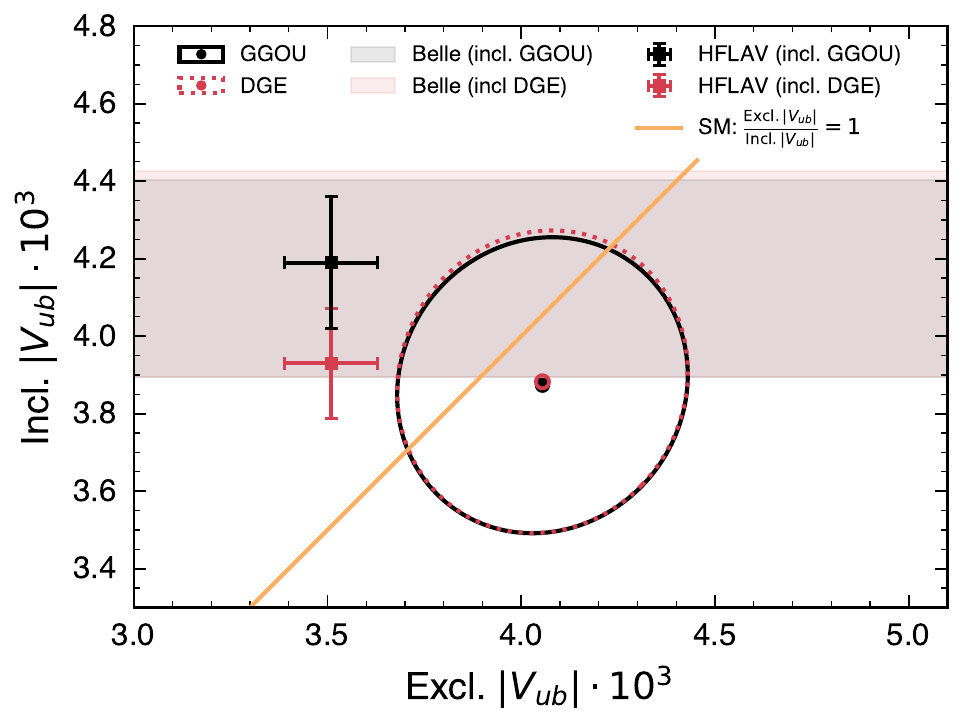} 
		  \includegraphics[width=0.49\linewidth]{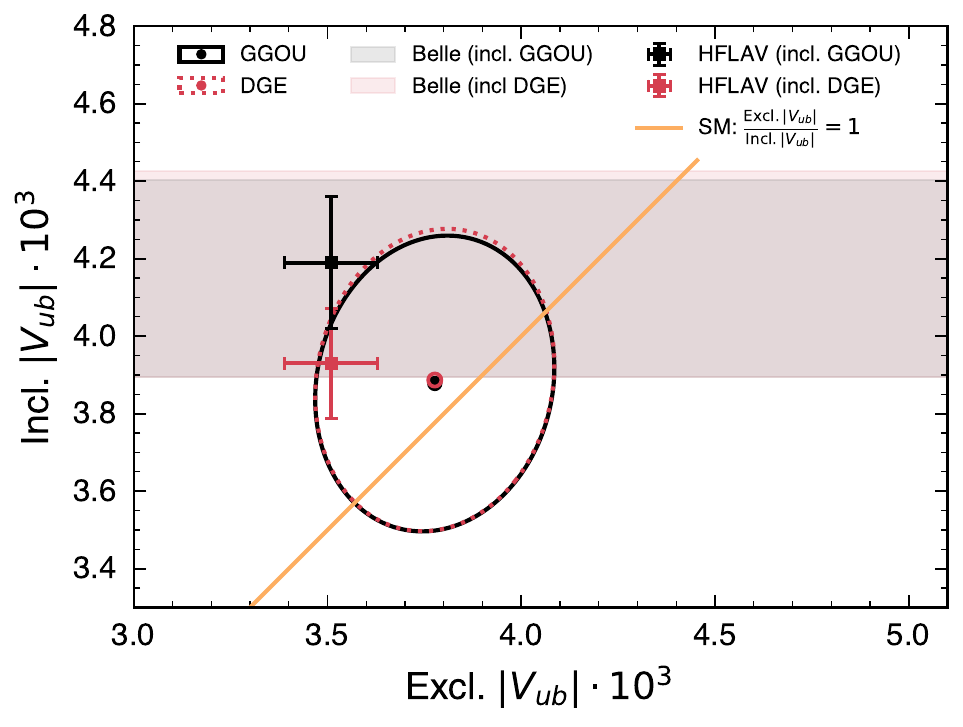} 
	\caption{ 
	The \AbsVub\ values obtained using the different theoretical inclusive decay rates are compared: GGOU versus BLNP (up) and GGOU versus DGE (low). The left column shows the fit with only LQCD constraints and the results from combined LQCD-experimental constraints are in the right column.
	}
\label{fig:others} 	
\end{figure}

Figure~\ref{fig:others} compares the inclusive \AbsVub values obtained from the GGOU calculation versus BLNP and DGE, respectively.  

\subsection{Data-MC agreement in background dominated sideband}

\begin{figure}[h!]
	\centering
		 \includegraphics[width=0.6\linewidth]{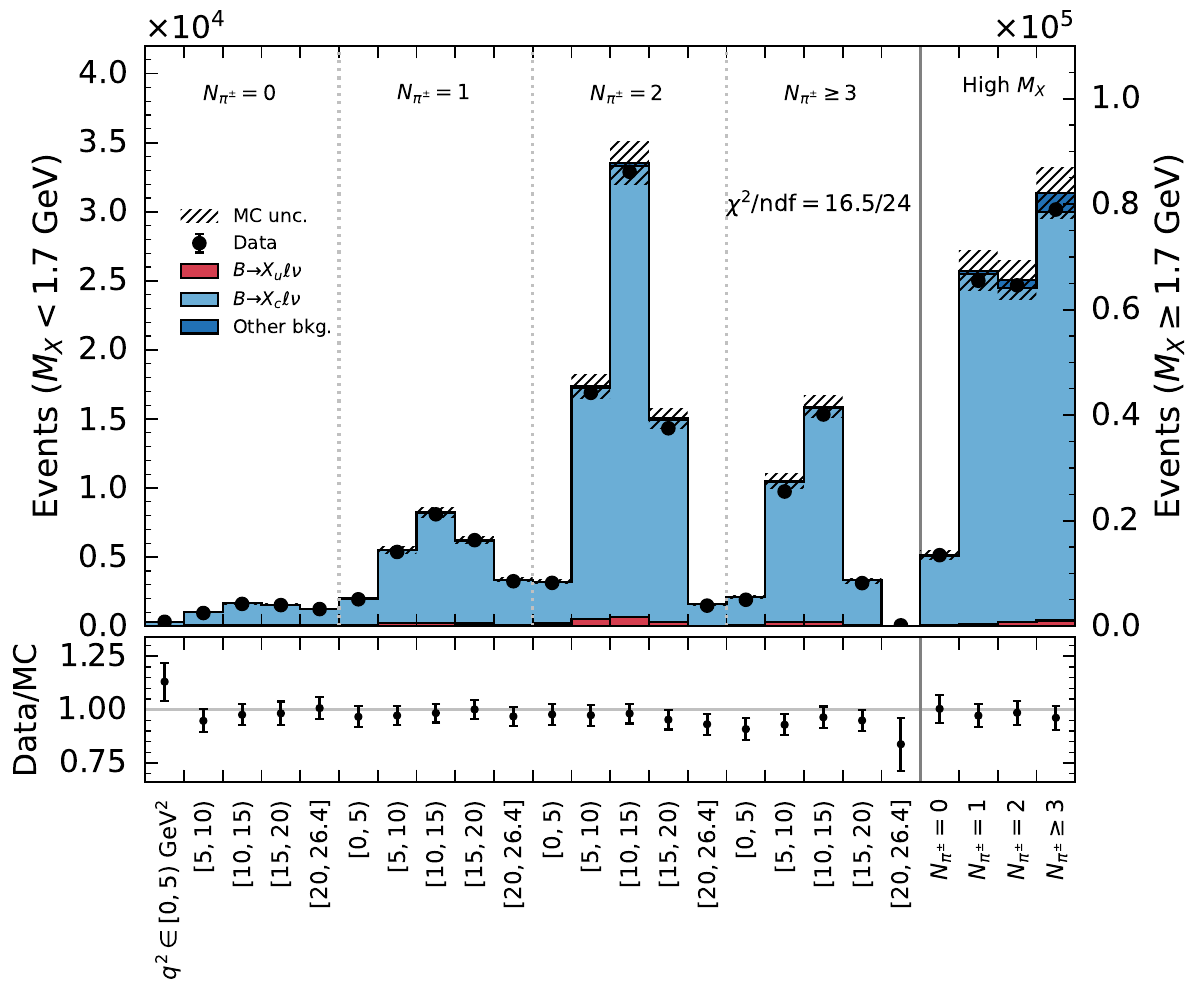} 	
	\caption{\label{fig:bkg} 
	The \bclnu sideband from the events rejected by the BDT selection is shown in the binning of the 2D fit. The uncertainties incorporate all systematic uncertainties discussed in the text. The bottom panel shows the ratio of observed events and the MC expectation. 
	}
\end{figure}

Figure~\ref{fig:bkg} shows the analyzed categories in $q^2:N_{\pi^\pm}$ for $M_X < 1.7 \, \mathrm{GeV}$ and the four $M_X \geq 1.7 \, \mathrm{GeV}$ bins in the \bclnu enriched BDT sideband. We observe fair agreement in the background shapes with a p-value of $87\%$.

\newpage
\subsection{Consistency cross-check}
The prediction is fitted to the observed data by minimizing
\begin{linenomath*}
\begin{align}
-2 \log \mathcal{L} &= -2 \log \prod_{i} \operatorname{Poisson}\left(\mathbf{\eta}_{\mathrm{obs}}, \mathbf{\eta}_{\mathrm{pred}}\cdot (1 + \epsilon \cdot \theta )\right)+\theta \rho_{\theta}^{-1} \theta^{T} + \chi^2_{\mathrm{FF}} \,, \\
\chi^2_{\mathrm{FF}} &= (\mathbf{a}_{\mathrm{obs}} - \mathbf{a}_{\mathrm{pred}}) \mathrm{Cov}^{-1}_{\mathrm{FF}}(\mathbf{a}_{\mathrm{obs}} - \mathbf{a}_{\mathrm{pred}})^{T} \,, 
\label{eq:fit} 
\end{align}
\end{linenomath*}
where the floating parameters $\mathbf{\eta}$ and $\mathbf{a}$ are the template normalization and the BCL form facotrs, respectively. The bin- and template-wise nuisance parameters $\theta$ are normalized to the relative bin errors $\epsilon$, and the associated bin-to-bin correlations arising from systematics are accounted in the fit by a global correlation matrix $\rho_{\theta}$. The BCL form factors are constrained by the covariance matrix $\mathrm{Cov}_{\mathrm{FF}}$ provided by FLAG.

In this measurement, an additional fit with separate normalizations of the \bpichlnu and \bpizlnu decays is applied to check the consistency of the nominal results based on combining the two modes. All of the fitter setups are summarised in the following:
\begin{itemize}
    \item Setup 1-a: fit $q^{2}:N_{\pi^{\pm}}$ spectra with LQCD and external experimental constraint on the BCL form factor and shared \bpilnu normalization based on the isospin relation.
    \item Setup 1-b: same as 1-a, but with only LQCD constraint for the form factor.
    \item Setup 2-a: separate normalizations of the \bpichlnu and \bpizlnu decays and with LQCD-experimental constraint. We denote the recovered CKM matrix values as $|V_{ub}|^{\pi^0 / \pi^+}$.
    \item Setup 2-b: same as 2-a, but with only LQCD constraint.
\end{itemize}

\begin{figure}
	\centering
		 \includegraphics[width=0.4\linewidth]{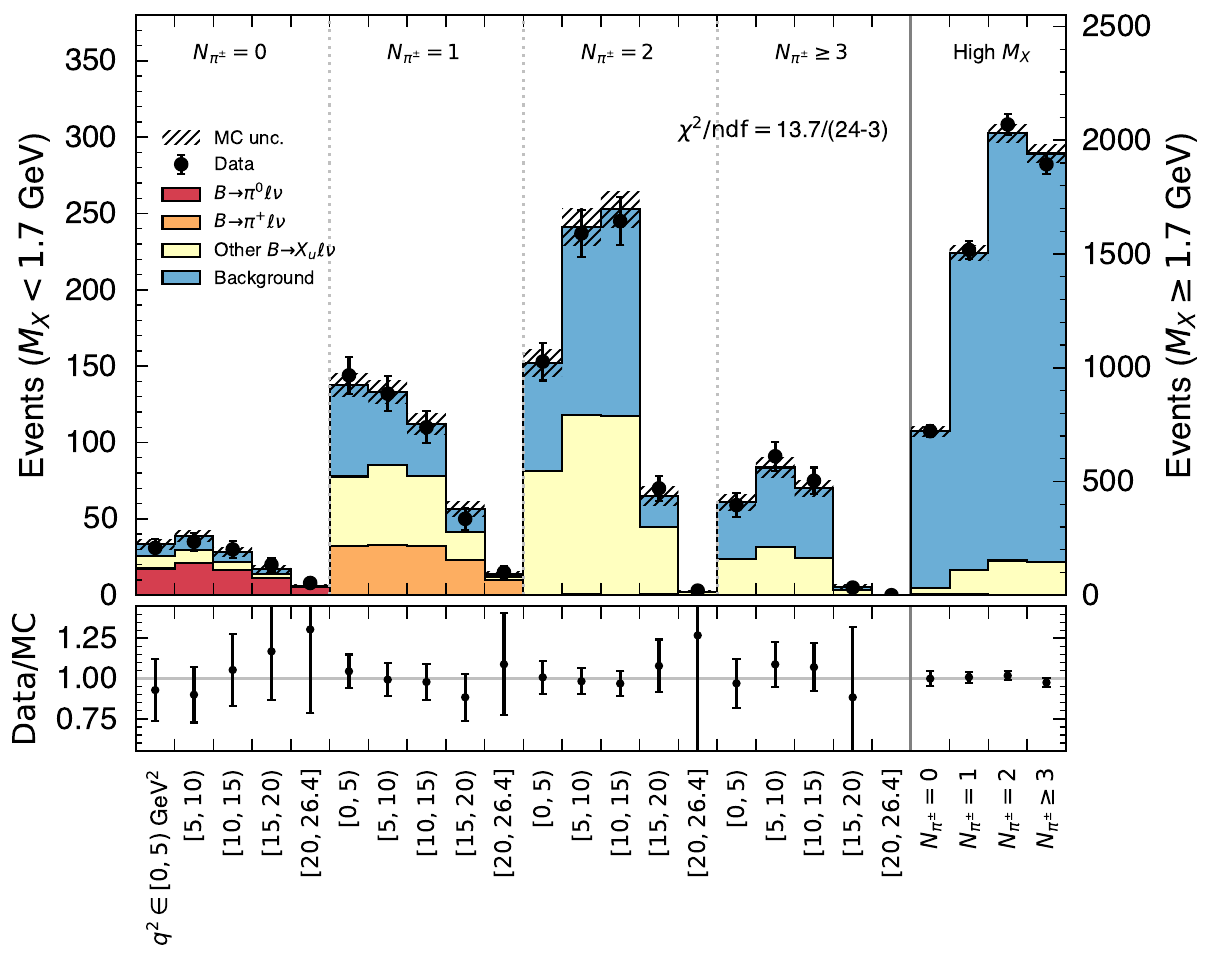}
            \includegraphics[width=0.4\linewidth]{figures/postfit_NLL_3mu_fit_FLAG.pdf} \\
            \includegraphics[width=0.4\linewidth]{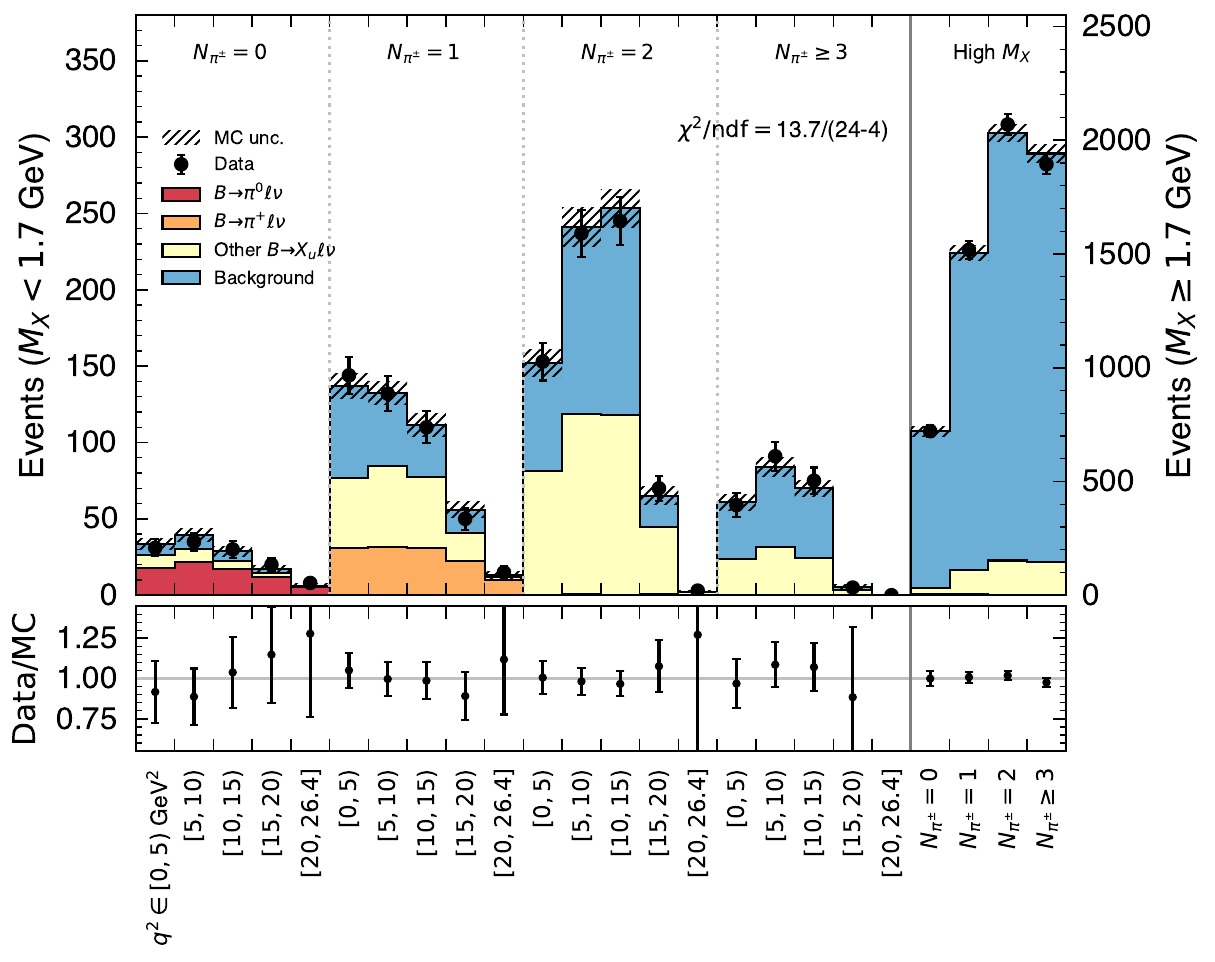} 
		 \includegraphics[width=0.4\linewidth]{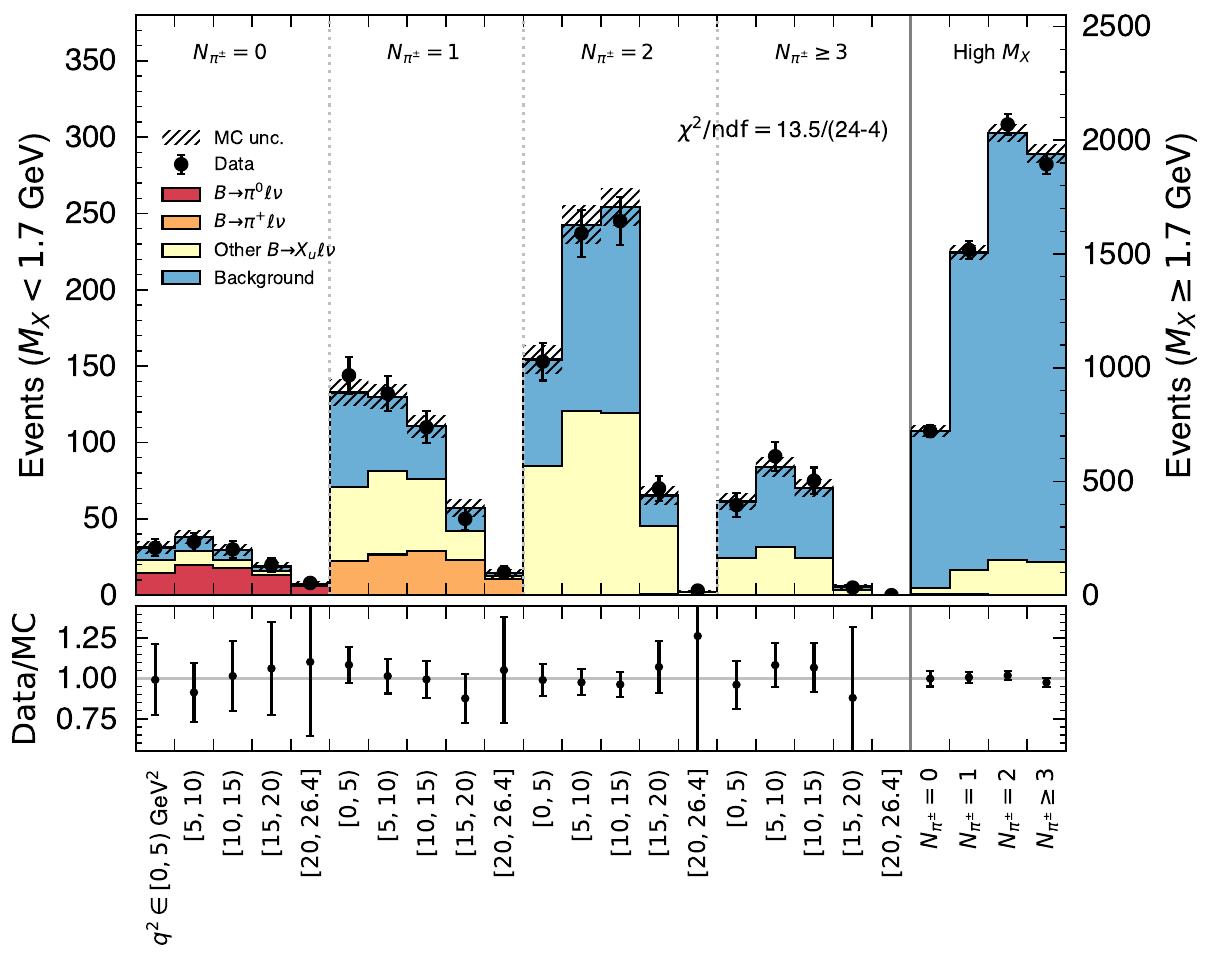}   
	\caption{ The postfit $q^{2}:N_{\pi^{\pm}}$ spectra with various setups.From top left to bottom right, the results are shown for the setup 1-a, 1-b, 2-a and 2-b. The uncertainties incorporate all post-fit uncertainties discussed in the main text.
	}
\label{fig:postfit-all} 	
\end{figure}

\begin{figure}
	\centering
          \includegraphics[width=0.24\linewidth]{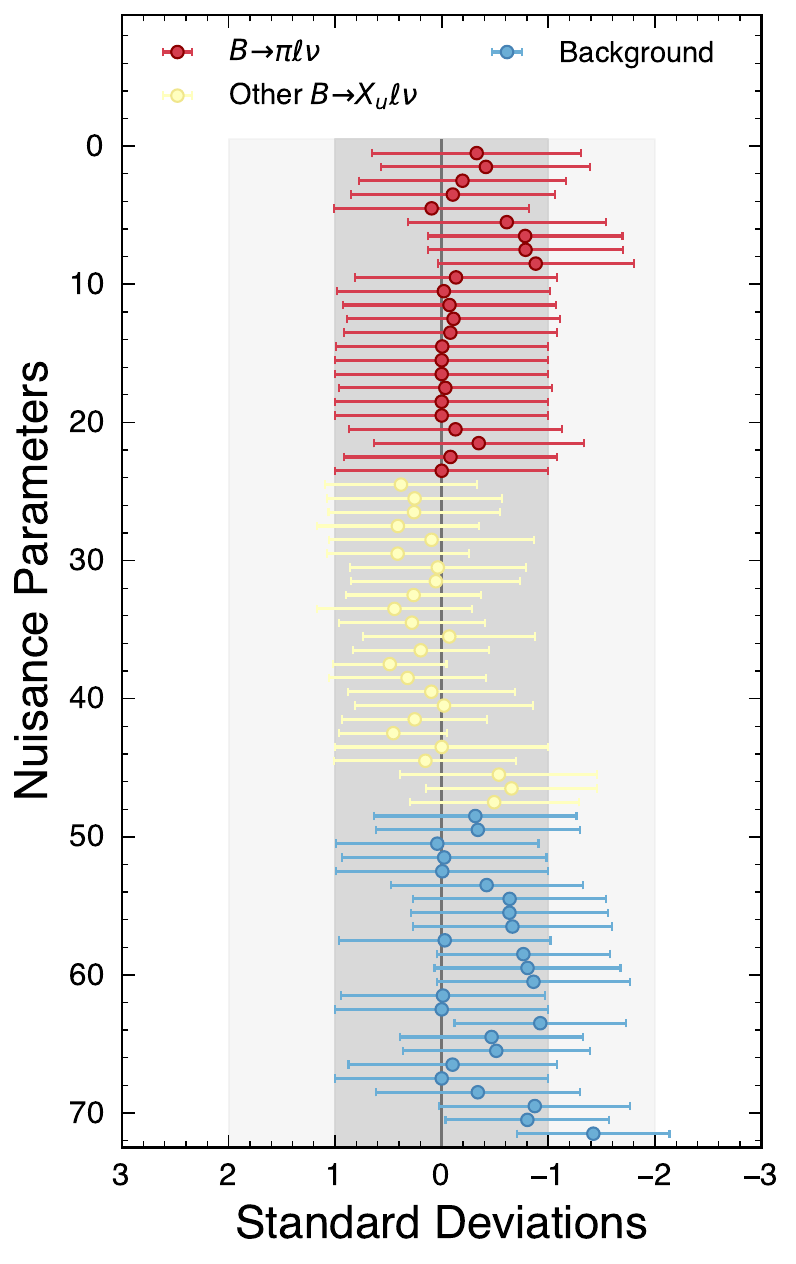} 
		 \includegraphics[width=0.24\linewidth]{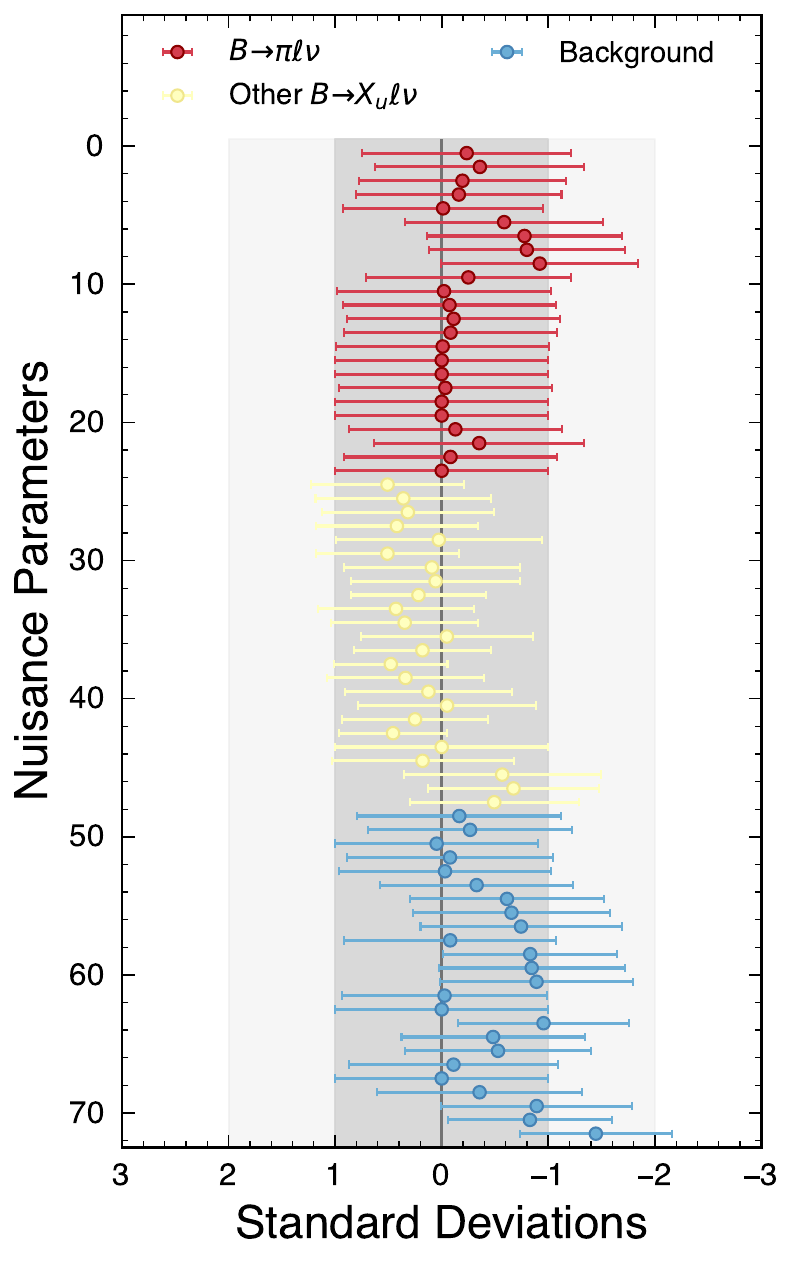} 
          \includegraphics[width=0.24\linewidth]{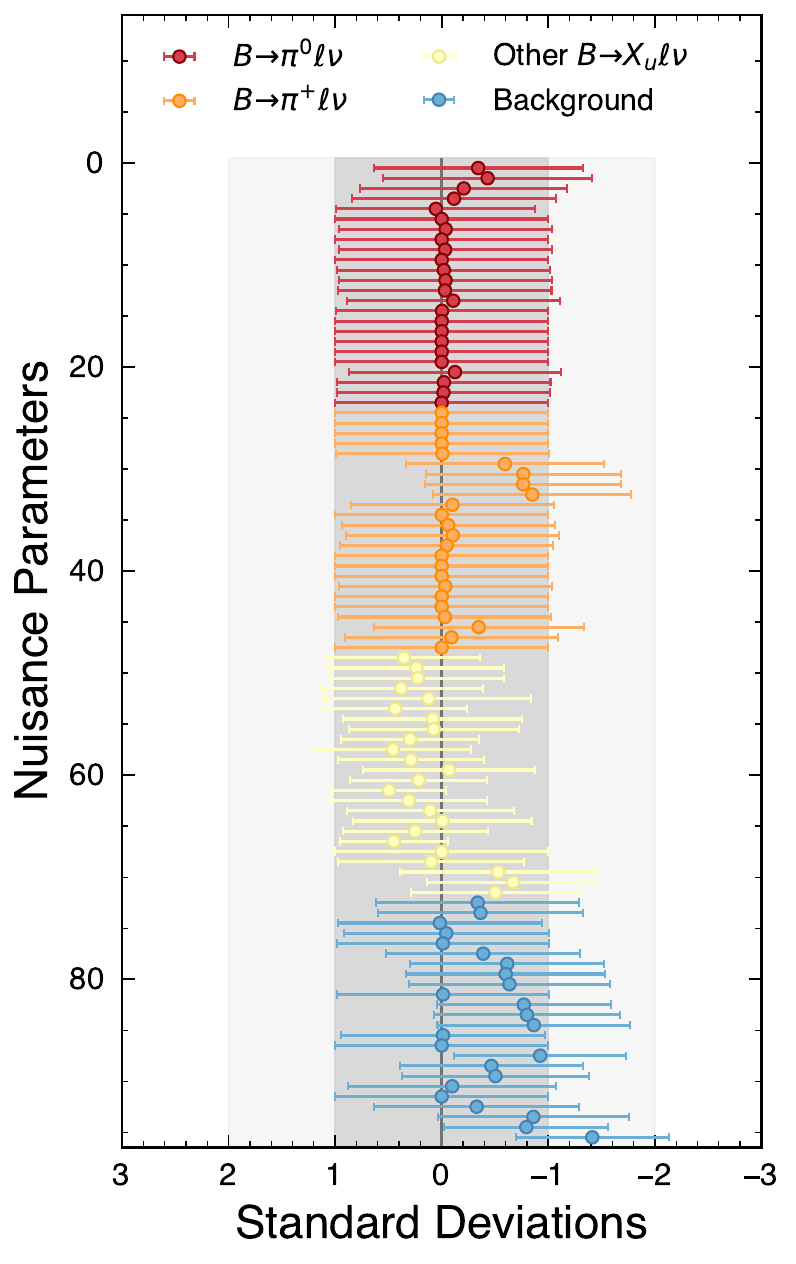} 
		 \includegraphics[width=0.24\linewidth]{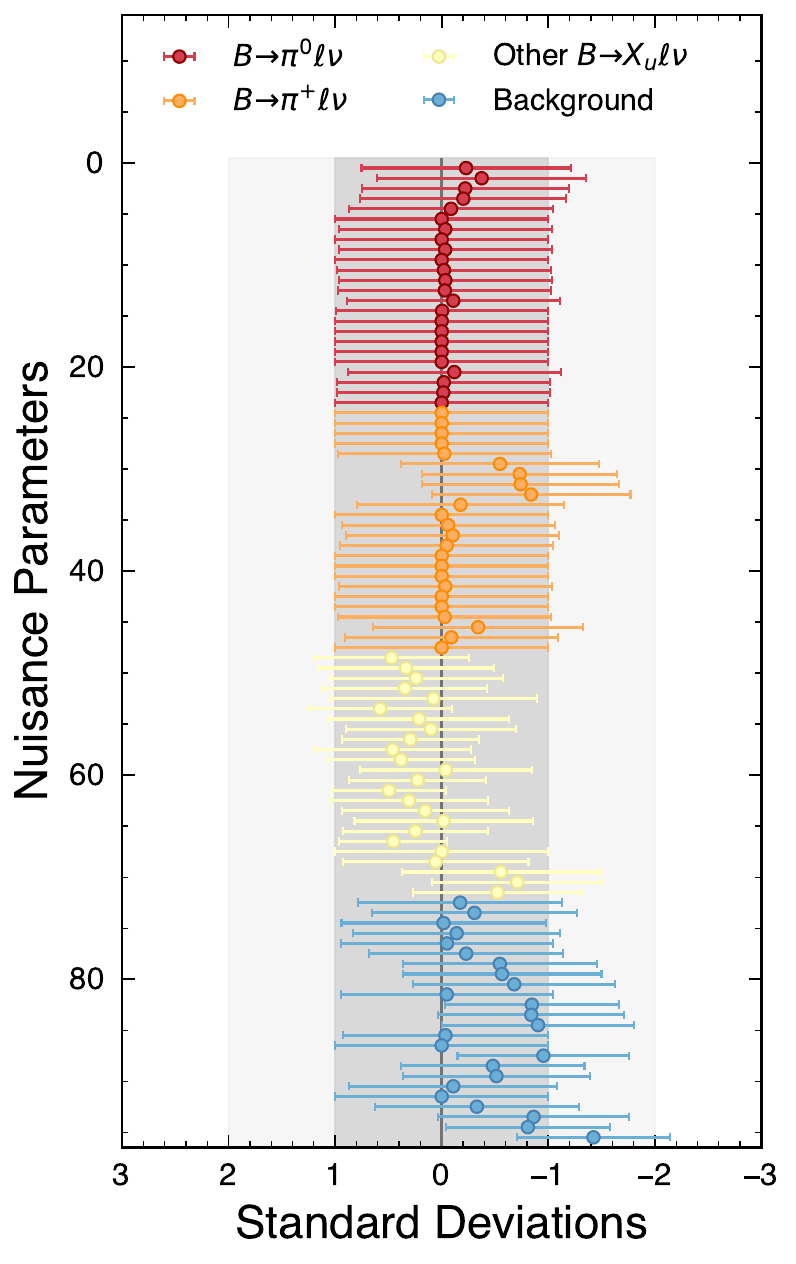} 	 
	\caption{ The pulls of bin-wise nuisance parameters. From left to right, the results are shown for the fit setup 1-a, 1-b, 2-a and 2-b. The uncertainty of each pull shows the post-fit error normalized to the pre-fit constraint.  
	}
\label{fig:pull-NPs} 	
\end{figure}

The nominal results are based on setup 1-a and 1-b. With different fit scenarios, the numerical results of the fitted yields are summarised in Table~\ref{tab:yields} as well as the signal efficiencies. After all selections, the total measured data are $7715 \pm 88$ events. Figure~\ref{fig:postfit-all} and Fig.~\ref{fig:pull-NPs} illustrate the post-fit spectra and the pulls of the template- and bin-wise Nuisance parameters, respectively. The featured behaviors are found to be consistent in all setups. The obtained \AbsVub and branching fractions are listed in Table~\ref{tab:vub-FLAG-4mu} and Table~\ref{tab:BR-FLAG-4mu} for the setup 2-a and 2-b, where the weighted average of two pion modes is derived based on the total covariance matrix. The final results are found to be fairly compatible with the nominal results in Table~\ref{tab:vub-FLAG-3mu} and Table~\ref{tab:BR-FLAG-3mu}.

\begin{table}[h!]
\renewcommand\arraystretch{1.2}
\begin{tabular}{lcccc}
 \hline \hline
Setup                  \;\;                     & \bpizlnu  \;\;\;\;    & \bpichlnu   \;\; \;\;    & Other \bulnu   \;\; \;\;       & Bkg.       \\
 \hline
1-a                                           & $75 \pm 11$ & $138 \pm 21$ & $1065 \pm 238$ & $6430 \pm 630$ \\

1-b                                             & $71 \pm 12 $ & $132 \pm 23 $ & $1076 \pm 238 $ & $6429 \pm 631 $      \\

2-a                                           & $78 \pm 14$ & $132 \pm 31$ & $1069 \pm 238$ & $6430 \pm 631$     \\

2-b                                               & $76 \pm 14$ & $117 \pm 33$ & $1087 \pm 238$ & $6428 \pm 631$     \\
\hline
$10^{3}\cdot\epsilon_{\mathrm{sig}}$          & $0.30$           & $0.31$      & $0.26$  & - \\ 
  \hline \hline 
\end{tabular}
\caption{The fitted yields for \bpizlnu, \bpichlnu, other \bulnu decays and backgrounds with various fitter setups. The uncertainties assigned to the fitted yields include the statistical and systematic impacts in the fitting procedure. The signal efficiencies $\epsilon_{\mathrm{sig}}$ are also listed.}
\label{tab:yields} 
\end{table}

\begin{table}
\renewcommand\arraystretch{1.2}
\centering
\begin{tabular}{lc}
\hline\hline
                 & Result \\
 \hline
 Setup 1-a &   \\
$|V_{ub}|^{\mathrm{excl.}}$                             & $(3.78 \pm 0.23 \pm 0.16 \pm 0.14)\times 10^{-3}$         \\

$|V_{ub}|^{\mathrm{incl.}}$                       & $(3.88 \pm 0.20 \pm 0.31 \pm 0.09)\times 10^{-3}$                  \\
$|V_{ub}|^{\mathrm{excl.}} / |V_{ub}|^{\mathrm{incl.}}$               & $0.97\pm 0.12$    \\  
$\rho(|V_{ub}|^{\mathrm{excl.}}, |V_{ub}|^{\mathrm{incl.}})$                &    $0.11$ \\  
 \hline
Setup 1-b &   \\

$|V_{ub}|^{\mathrm{excl.}}$                             & $(4.05 \pm 0.30 \pm 0.16 \pm 0.16)\times 10^{-3}$         \\

$|V_{ub}|^{\mathrm{incl.}}$                       & $(3.87 \pm 0.20 \pm 0.31 \pm 0.09)\times 10^{-3}$                  \\
$|V_{ub}|^{\mathrm{excl.}} / |V_{ub}|^{\mathrm{incl.}}$               & $1.05\pm 0.14$    \\  
$\rho(|V_{ub}|^{\mathrm{excl.}}, |V_{ub}|^{\mathrm{incl.}})$                &    $0.07$ \\    
\hline\hline
\end{tabular}
\caption{The determined \AbsVub results and various ratios with the setup 1-a and 1-b, respectively.}
\label{tab:vub-FLAG-3mu} 
\end{table}

\begin{table}
\renewcommand\arraystretch{1.2}
\centering
\begin{tabular}{lc}
\hline\hline
                 & Result \\
 \hline
Setup 1-a &   \\
$\mathcal{B}(\bpichlnu)$                           & $(1.53\pm 0.18 \pm 0.12 )\times 10^{-4}$                  \\
$\Delta \mathcal{B}(\bulnu)$                       & $(1.39 \pm 0.14 \pm 0.22)\times 10^{-3}$                  \\
$\rho(\mathcal{B}^{\pi}, \Delta \mathcal{B}^{X_{u}})$                &    $0.12$ \\   
 \hline
Setup 1-b &   \\

$\mathcal{B}(\bpichlnu)$                             & $(1.45 \pm 0.19 \pm 0.14 )\times 10^{-4}$         \\

$\Delta \mathcal{B}(\bulnu)$                       & $(1.39 \pm 0.14 \pm 0.22)\times 10^{-3}$                  \\
$\rho(\mathcal{B}^{\pi}, \Delta \mathcal{B}^{X_{u}})$                &    $0.11$ \\  
\hline\hline
\end{tabular}
\caption{The measured branching fractions and various correlations based on the setup 1-a and 1-b, respectively.}
\label{tab:BR-FLAG-3mu} 
\end{table}

\begin{table}
\renewcommand\arraystretch{1.2}
\centering
\begin{tabular}{lc}
\hline\hline
                 & Result \\
 \hline
 Setup 2-a &   \\
$|V_{ub}|^{\pi^{0}}$                             & $(3.86 \pm 0.30 \pm 0.18 \pm 0.15)\times 10^{-3}$         \\
$|V_{ub}|^{\pi^{+}}$                            & $(3.69\pm 0.34 \pm 0.24 \pm 0.10)\times 10^{-3}$              \\
Avr. $|V_{ub}|^{\mathrm{excl.}}$                            & $(3.79\pm 0.31)\times 10^{-3}$              \\
$|V_{ub}|^{\mathrm{incl.}}$                       & $(3.88 \pm 0.20 \pm 0.31 \pm 0.09)\times 10^{-3}$                  \\
$|V_{ub}|^{\mathrm{excl.}} / |V_{ub}|^{\mathrm{incl.}}$               & $0.98\pm 0.12$    \\  
$\rho(|V_{ub}|^{\mathrm{excl.}}, |V_{ub}|^{\mathrm{incl.}})$                &    $0.10$ \\    
$\rho(|V_{ub}|^{\pi^{+}}, |V_{ub}|^{\pi^{0}})$                &    $0.20$ \\   
 \hline
Setup 2-b &   \\
$|V_{ub}|^{\pi^{0}}$                             & $(4.31 \pm 0.44 \pm 0.25 \pm 0.16)\times 10^{-3}$         \\
$|V_{ub}|^{\pi^{+}}$                            & $(3.88\pm 0.37 \pm 0.23 \pm 0.14)\times 10^{-3}$              \\
Avr. $|V_{ub}|^{\mathrm{excl.}}$                            & $(4.06\pm 0.38)\times 10^{-3}$              \\
$|V_{ub}|^{\mathrm{incl.}}$                       & $(3.87 \pm 0.20 \pm 0.32 \pm 0.09)\times 10^{-3}$                  \\
$|V_{ub}|^{\mathrm{excl.}} / |V_{ub}|^{\mathrm{incl.}}$               & $1.05\pm 0.14$    \\  
$\rho(|V_{ub}|^{\mathrm{excl.}}, |V_{ub}|^{\mathrm{incl.}})$                &    $0.06$ \\    
$\rho(|V_{ub}|^{\pi^{+}}, |V_{ub}|^{\pi^{0}})$                &    $0.22$ \\   
\hline\hline
\end{tabular}
\caption{The determined \AbsVub results and various ratios based on the setup 2-a and 2-b, respectively.}
\label{tab:vub-FLAG-4mu} 
\end{table}

\begin{table}
\renewcommand\arraystretch{1.2}
\centering
\begin{tabular}{lc}
\hline\hline
                 & Result \\
 \hline
Setup 2-a &   \\
$\mathcal{B}(\bpizlnu)$                             & $(0.85 \pm 0.13 \pm 0.08 )\times 10^{-4}$         \\
$\mathcal{B}(\bpichlnu)$                           & $(1.46\pm 0.26 \pm 0.19 )\times 10^{-4}$                  \\
$\Delta \mathcal{B}(\bulnu)$                       & $(1.39 \pm 0.15 \pm 0.22)\times 10^{-3}$                  \\
$\rho(\mathcal{B}^{\pi^{0}}, \mathcal{B}^{\pi^{+}})$                &    $0.08$ \\    
$\rho(\mathcal{B}^{\pi^{0}}, \Delta \mathcal{B}^{X_{u}})$                &    $0.08$ \\  
$\rho(\mathcal{B}^{\pi^{+}}, \Delta \mathcal{B}^{X_{u}})$                &    $0.08$ \\   
 \hline
Setup 2-b &   \\

$\mathcal{B}(\bpizlnu)$                             & $(0.84 \pm 0.13 \pm 0.08 )\times 10^{-4}$         \\
$\mathcal{B}(\bpichlnu)$                           & $(1.27\pm 0.27 \pm 0.20 )\times 10^{-4}$                  \\
$\Delta \mathcal{B}(\bulnu)$                       & $(1.38 \pm 0.14 \pm 0.22)\times 10^{-3}$                  \\
$\rho(\mathcal{B}^{\pi^{0}}, \mathcal{B}^{\pi^{+}})$                &    $0.14$ \\    
$\rho(\mathcal{B}^{\pi^{0}}, \Delta \mathcal{B}^{X_{u}})$                &    $0.08$ \\  
$\rho(\mathcal{B}^{\pi^{+}}, \Delta \mathcal{B}^{X_{u}})$                &    $0.06$ \\   
\hline\hline
\end{tabular}
\caption{The measured branching fractions and various correlations based on the setup 2-a and 2-b, respectively.}
\label{tab:BR-FLAG-4mu} 
\end{table}

\subsection{BCL parameters of \bpilnu decay form factor}
\begin{figure}[h]
	\centering
        \includegraphics[width=0.48\linewidth]{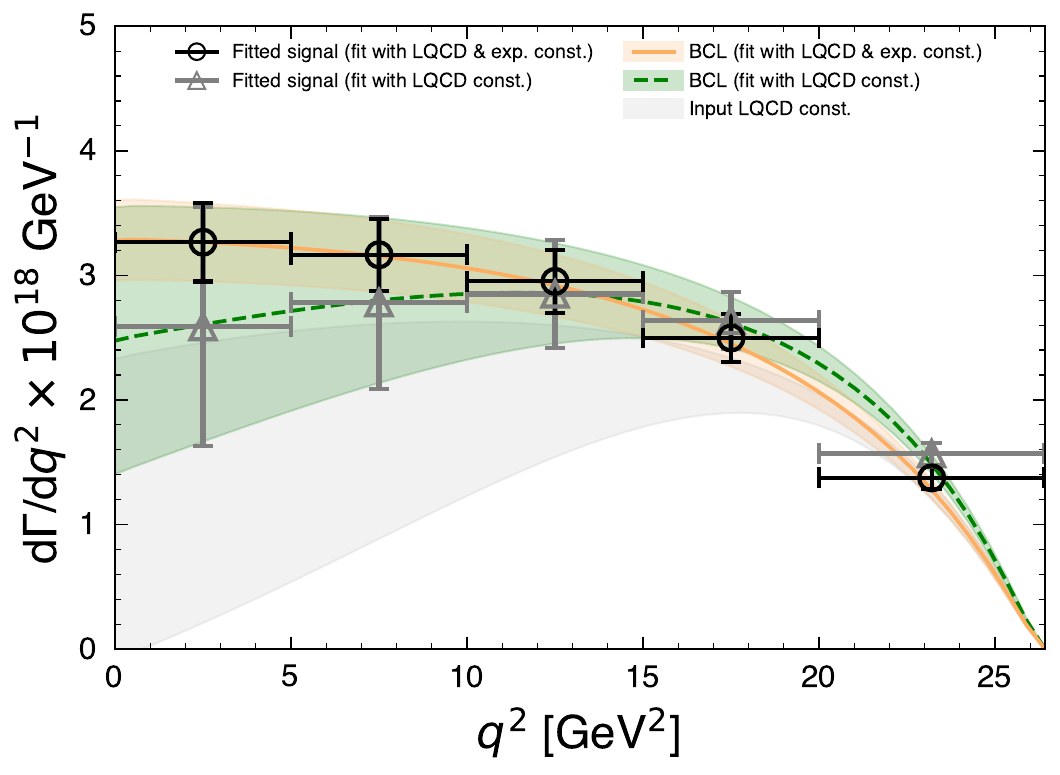} \\
		 \includegraphics[width=0.48\linewidth]{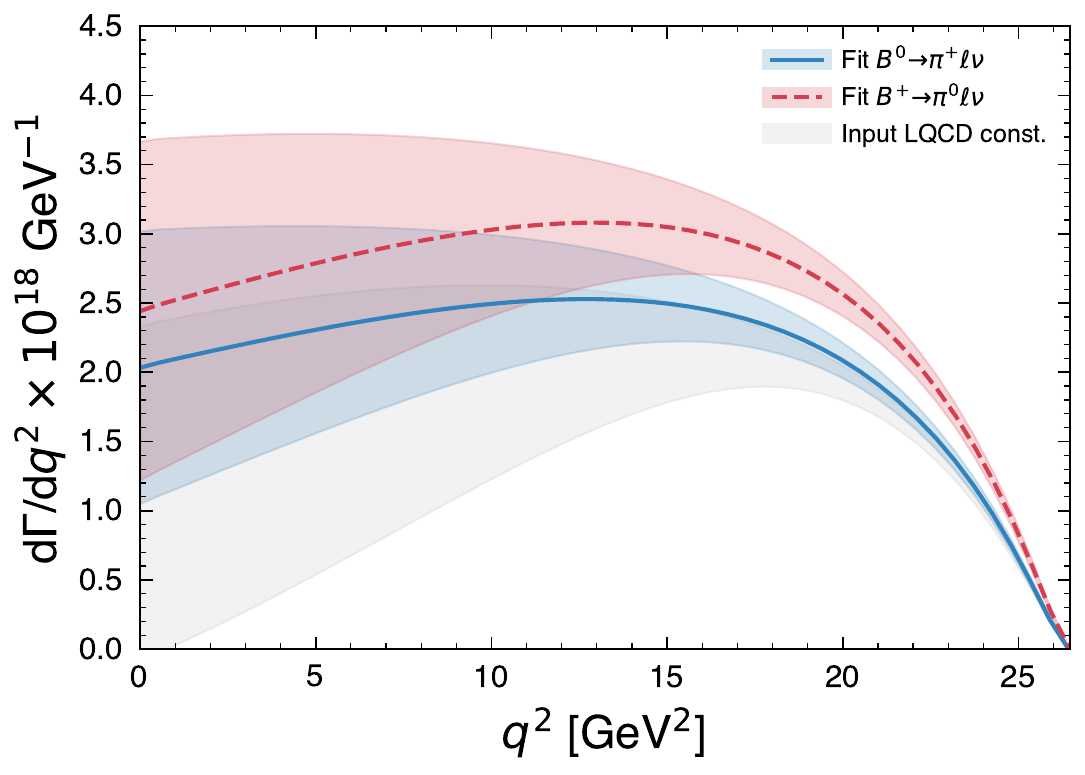} 
		 \includegraphics[width=0.48\linewidth]{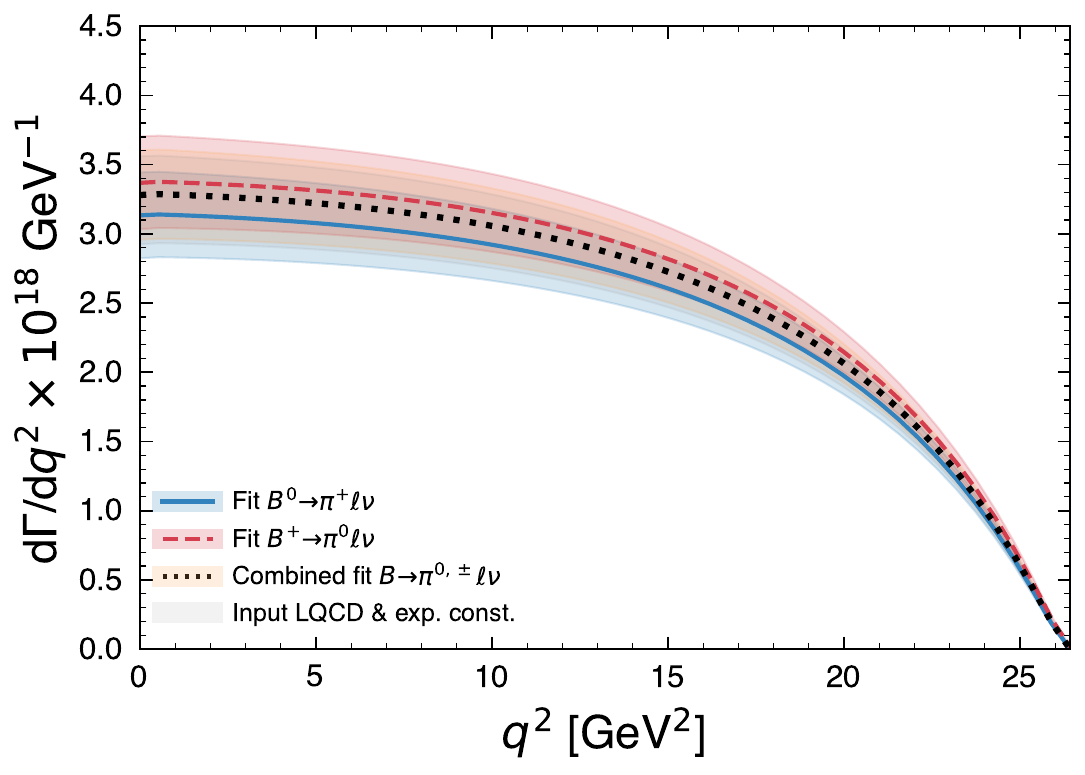} 
	\caption{ Top: the $q^{2}$ spectra of \bpilnu obtained from the fit of the combined LQCD and experimental information (orange, solid) and from the fit to LQCD only (green, dashed) are shown. The data points are the post-fit signal distributions, corrected for resolution and efficiency effects and averaged over both isospin modes. The input LQCD constraints from FLAG are shown in grey.
	Bottom left: the $q^{2}$ spectra obtained with separated (blue, solid) $\pi^{+}$ mode and (red, dashed) $\pi^{0}$ using the LQCD only information from FLAG to constrain the \bpilnu form factor (setup 2-b). Bottom right: the results obtained by using the LQCD and experimental constraint (setup 2-a). The combined fit (setup 1-a) result is shown for comparison (black, dotted).
	}
\label{fig:BCL-FLAG-suppl} 	
\end{figure}

The fitted BCL parameters are summarized in Table~\ref{tab:FF-FLAG} and \ref{tab:FF-FLAG-wExp} with only LQCD constraints and combined LQCD-experimental constraints, respectively. Figure~\ref{fig:BCL-FLAG-suppl} compares the results obtained in various fit scenarios, which are in good agreement. 

\begin{table}[h]
\renewcommand\arraystretch{1.2}
\centering
\begin{tabular}{lrrrrrr}
\hline\hline
&  \AbsVub $\times 10^{3}$ &    $a^{+}_{0}$ &   $a^{+}_{1}$ &    $a^{+}_{2}$ &     $a^{0}_{0}$ &  $a^{0}_{1}$ \\
\hline
Central &   4.055 &  0.407 & -0.597 & -0.465 &  0.496 & -1.504 \\
Uncertainty & 0.375 &  0.012 &  0.080 &  0.392 &  0.020 &  0.096 \\
\hline
\AbsVub & 1.000 & -0.416 & -0.473 & -0.308 & -0.184 & -0.462 \\
$a^{+}_{0}$ &    &   1.000 &  0.275 & -0.187 &  0.254 &  0.173 \\
$a^{+}_{1}$ &   &    &   1.000 &  0.344 &  0.101 &  0.720 \\
$a^{+}_{2}$ &    &   &    &   1.000 &  0.193 &  0.698 \\
$a^{0}_{0}$ &   &    &    &    & 1.000 & -0.039 \\
$a^{0}_{1}$ &    &   &    &   &   &  1.000 \\
\hline\hline
\end{tabular}
\caption{The measured $B\to \pi \ell \nu$ form factor BCL parameters and exclusive \AbsVub\ with full correlations. The shape of $q^{2}$ is constrained by the LQCD fit results from FLAG.}
\label{tab:FF-FLAG} 
\end{table}

\begin{table}[h]
\renewcommand\arraystretch{1.2}
\centering
\begin{tabular}{lrrrrrr}
\hline\hline
&  \AbsVub $\times 10^{3}$ &    $a^{+}_{0}$ &   $a^{+}_{1}$ &    $a^{+}_{2}$ &     $a^{0}_{0}$ &  $a^{0}_{1}$ \\
\hline
Central &    3.777 &  0.414 & -0.493 & -0.297 &  0.500 & -1.426 \\
Uncertainty &  0.309 &  0.014 &  0.053 &  0.180 &  0.023 &  0.054 \\
\hline
\AbsVub &  1.000 & -0.452 & -0.168 &  0.232 & -0.109 & -0.105 \\
$a^{+}_{0}$ &    &   1.000 &  0.151 & -0.451 &  0.259 &  0.142 \\
$a^{+}_{1}$ &   &    &   1.000 & -0.798 & -0.096 &  0.214 \\
$a^{+}_{2}$ &    &   &   & 1.000 &  0.012 & -0.097 \\
$a^{0}_{0}$ &    &   &  &   &  1.000 & -0.451 \\
$a^{0}_{1}$ &   &  &    &   &   &  1.000 \\
\hline\hline
\end{tabular}
\caption{The measured $B\to \pi \ell \nu$ form factor BCL parameters and exclusive \AbsVub\ with full correlations. The shape of $q^{2}$ is constrained by the combined LQCD and experimental fit results from FLAG.}
\label{tab:FF-FLAG-wExp} 
\end{table}

\end{document}